# Investigation of the Pressure Dependent Physical Properties of MAX Phase Ti$_2$Al$X$ ($X$ = B, C, and N) Compounds: A First-Principles Study

M.I. Naher[1], M. Montasir[2], M.Y.H. Khan[2], M.A. Ali[3,4], M.M. Hossain[3,4], M.M. Uddin[3,4], M.Z. Hasan[2], M.A. Hadi[5], S.H. Naqib[4,5*]

[1]Department of Mathematics and Natural Sciences, BRAC University, Dhaka-1212, Bangladesh

[2]Department of Electrical and Electronic Engineering, International Islamic University Chittagong, Kumira, Chittagong, 4318, Bangladesh

[3]Department of Physics, Chittagong University of Engineering and Technology (CUET), Chattogram-4349, Bangladesh

[4]Advanced Computational Materials Research Laboratory, Department of physics, Chittagong University of Engineering and Technology (CUET), Chattogram-4349, Bangladesh

[5]Department of Physics, University of Rajshahi, Rajshahi 6205, Bangladesh

*Corresponding author; Email: salehnaqib@yahoo.com

**Abstract**

The physical properties and their pressure dependence of recently synthesized Ti$_2$Al$X$ ($X$ = B, C, and N) MAX phases are investigated for the very first-time applying density functional theory (DFT). The optimized lattice parameters are in good agreement with the existing literature. The metallic character of the phases is supported by the Fermi level overlap of the conduction band and valence band. All three compounds are structurally, mechanically, and dynamically stable at ambient pressure and at pressures up to 10 GPa. The materials are brittle in nature at 0 GPa and start exhibiting ductility at around 8 GPa, confirmed by both Pugh's and Poisson's ratios. All three compounds possess good level of hardness. The compounds have moderate machinability. The Debye temperature, phonon thermal conductivity, and melting temperature are moderate; with the highest value observed for Ti$_2$AlN. The lattice thermal conductivity decreases with increasing pressure (up to 10 GPa) and temperature (up to 1500 K). Directional anisotropies in mechanical, optical, and thermal features are at moderate level for all three MAX phases. All the compounds under study exhibit potential as promising TBC (thermal barrier coating) materials. They are also suitable for optical applications.

## 1. Introduction

The MAX phases are considered as a class of solids with the chemical formula M$_{n+1}$AX$_n$ (n = 1-3) belonging to a family of ternary borides, carbides, and nitrides. Here, "M" is an early transition metal, "A" is an A-group element (mainly III A and IV A), and "X" is either the element B or C or N [1-3]. The MAX phase is classified into three groups based on the n value, viz. 211, 312, and 413. MAX compounds feature an excellent set of properties of both metal and ceramic, which bridges the metal-ceramic gap. The growing interest in the MAX phase over the last few decades is attributed to their alluring properties, including light weight (low density), high melting point, exceptionally damage tolerant behavior at room temperature, excellent conductivity (both thermal and electrical), resistance to thermal shock and chemical attack, high fracture toughness, plastic and high oxidation resistance at high-temperature conditions, self-healing cracks, relatively low thermal expansion coefficient (TEC) of $< 10 \times 10^{-6}$ ºC$^{-1}$, radiation tolerance, and easy machinability [2-8]. These features make MAX phase materials compatible with a wide range of industrial and technological applications such as in Li-ion batteries, sliding electrical contacts, wear and corrosion-resistant coatings, superconducting materials, spintronics, and the nuclear shielding sector [9,10].

Ti-based 211 MAX phase compounds such as Ti$_2$AlB, Ti$_2$AlC, and Ti$_2$AlN are reported to have significant potential for applications. Among those, Ti$_2$AlB is a theoretically predicted new phase [9,11] that has not yet been experimentally confirmed. Some physical properties of Ti$_2$AlB, such as mechanical, electronic, acoustic velocities, Debye temperature, minimum thermal conductivity, and phonon dynamics have been studied at ambient and only at a rudimentary level [9,11]. However, Ti$_2$AlC and Ti$_2$AlN have already been experimentally synthesized [7,12,13], and some theoretical studies have been reported [11,14-17]. Ti$_2$AlC has already attracted attention from the industry due to its exotic behavior such as it is reported to exhibit excellent oxidation resistance [18], good TBC [3] potential, brittle-to-ductile transition at 1050 °C [18], sufficient damage tolerance to irradiations [19], high yield points and significant plasticity at temperatures as high as 1300°C in air [6]. The TEC of Ti$_2$AlC and Ti$_2$AlN are within $8.2 \times 10^{-6}$ and $8.8 \times 10^{-6}$ ºC$^{-1}$, respectively, in the temperature range from 25 ºC to 1300 ºC [5]. The electrical conductivity of Ti$_2$AlC and Ti$_2$AlN are $2.7 \times 10^6$ and $4 \times 10^6$ (Ωm)$^{-1}$, respectively, at room temperature [5, 20]. Some other physical properties of Ti$_2$AlC and Ti$_2$AlN have also been studied employing *ab-initio* calculations in the ground state [11,14,16,20-22]. Overall, only a few studies on Ti$_2$Al$X$ ($X$ = B, C, and N) materials have been conducted so far, both experimentally and theoretically [5,11]. Surprisingly, a combination of physical properties and their responses to pressure and temperature have not been comprehensively explored, either experimentally or theoretically. A systematic study of these materials is therefore necessary. Thus, the prime objective of this article is to explore the physical properties of the Ti-based MAX phase Ti$_2$Al$X$ ($X$ = B, C, and N) extensively under hydrostatic pressure up to 10 GPa by employing the density functional theory (DFT). Pressure is a clean tuning parameter to alter the physical properties of solids. We have increased the pressure from 0 GPa, in 2 GPa increments, up to 10 GPa. We also studied the effect of temperature on lattice thermal conductivity.

The manipulation of solids through the application of pressure, temperature, electric or magnetic fields, and variations in composition via doping and substitution can bring about significant modifications in the physical properties of these materials. This is particularly important as certain materials require changes in their properties to enhance their suitability for commercial applications. A recent area of research interest involves the investigation of the effects of applying pressure specifically on MAX phases [23-25]. The general response of MAX phase materials to pressure includes structural, mechanical, electronic, superconducting, and magnetic transitions, as well as increase in the superconducting transition temperature; all of which occurs at specific pressures. Additionally, many other physical parameters such as density, hardness (chemical bonding), machinability, acoustic velocities, melting temperature, Debye temperature, and thermal conductivity vary with pressure. This research, thus, has the potential to uncover new insights and applications for these materials.

The rest of this manuscript proceeds along the following outline: Section 2 contains a brief description of the computational methodology. Section 3 includes an in-depth investigation of the physical properties of Ti$_2$Al$X$ ($X$ = B, C, and N) along with their possible implications. Finally, in Section 4, we summarized the key findings and draw conclusions from this work.

## 2. Computational scheme

The first-principles calculations were carried out using the density functional theory [26,27] as contained in the CAmbridge Serial Total Energy Package (CASTEP) code [28]. The electronic valence states for Ti, Al, B, C, and N elements are taken as $3s^2\ 3p^6\ 3d^2\ 4s^2$, $3s^2\ 3p^1$, $2s^2\ 2p^1$, $2s^2\ 2p^2$, and $2s^2\ 2p^3$, respectively. The exchange–correlation energy was evaluated using the functional called generalized gradient approximation (GGA) incorporating the Perdew-Burke-Ernzerhof for solids (PBEsol) scheme [29]. The interaction between the valence electrons and ion cores was modeled with the Vanderbilt-type ultrasoft pseudopotential [30]. The optimized (minimum total energy and internal forces) structure of the studied MAX phases was found by applying the Limited-memory Broyden–Fletcher–Goldfarb–Shanno (LBFGS) minimization algorithm [31]. The plane-wave cut-off energy was set at 550 eV for Ti$_2$Al$X$ ($X$ = B, C, and N). The irreducible Brillouin zone was sampled with a 16×16×4 $k$-grid on the Monkhorst–Pack scheme [32]. Various tolerance parameters, during computations for all three compounds, were set to be 5×10$^{-6}$ eV/atom for energy, maximum lattice point displacement within 5×10$^{-4}$ Å, maximum ionic force within 10$^{-2}$ eVÅ$^{-1}$, maximum stress within 0.02 GPa and smearing width of 0.1 eV. A $k$-point mesh of 24×24×6 was used for the Fermi surface construction since it demands denser $k$-mesh. Moreover, the density functional perturbation theory (DFPT) finite displacement method (FDM) [33,34] has been employed to calculate phonon dispersion spectra (PDS) and total phonon density of states (PHDOS). All the DFT simulations were performed at the default temperature (0 K) and at different hydrostatic pressures in GPa.

The second-order elastic constant ($C_{ij}$) of the studied compounds were obtained employing the stress–strain method [35]. In the case of the hexagonal structure, there are six independent elastic stiffness constants: $C_{11}$, $C_{33}$, $C_{44}$, $C_{12}$, $C_{66}$, and $C_{13}$. Lattice dynamical properties such as phonon dispersion and phonon density of states were calculated with a conventional cell of each system using the finite displacement method. In these calculations, $k$-point mesh of 9×9×2, 9×9×2 (10×10×2), and 10×10×2 (8×8×2) was used for Ti$_2$AlB, Ti$_2$AlC, and Ti$_2$AlN at 0 GPa (10 GPa). A cut-off energy of 350 eV was set for all three compounds.

The frequency-dependent optical spectra are derived from their complex dielectric function given by $\varepsilon(\omega) = \varepsilon_1(\omega) + i\varepsilon_2(\omega)$. The imaginary part $\varepsilon_2(\omega)$ has been obtained from the following formula:

$$\varepsilon_2(\omega) = \frac{2e^2\pi}{\Omega\varepsilon_0} \sum_{k,v,c} |\langle \Psi_k^c|\hat{u}.\vec{r}|\Psi_k^v\rangle|^2\ \delta(E_k^c - E_k^v - E) \tag{1}$$

here, ω is the frequency of the incident photon, Ω is the unit cell volume, e is the electronic charge, $\varepsilon_0$ defines the permittivity of the free space, $\hat{u}$ gives the polarization of the incident electric field, and $\vec{r}$ is the position vector. Besides, $\Psi_k^c$ and $\Psi_k^v$ refer to the conduction and valence band wave functions at a given wave-vector $k$, respectively. This formula needs inputs from the electronic band structure calculations. With the help of Kramers-Kronig relationship, the real part $\varepsilon_1(\omega)$ has been evaluated from its corresponding imaginary part $\varepsilon_2(\omega)$ as [36]:

$$\varepsilon_1(\omega) = 1 + \frac{2}{\pi} P \int_0^\infty \frac{\omega' \varepsilon_2(\omega')}{\omega'^2 - \omega^2} d\omega' \qquad (2)$$

here, $P$ signifies the principal value of the integral.

Once $\varepsilon(\omega)$ is known, all the other optical parameters, such as refractive index $N(\omega)$, reflectivity $R(\omega)$, optical conductivity $\sigma(\omega)$, absorption coefficient $\alpha(\omega)$, and energy loss-function $L(\omega)$ can be deduced from it using standard formulae [37,38].

## 3. Results and analysis
### 3.1. Optimized structure and pressure dependency

Fig. 1 displays the optimized unit cell structure of $Ti_2AlX$ ($X$ = B, C, and N). An edge-shared octahedral of Ti$X$ is sandwiched with Al atomic sheets. Like all the MAX phases, $Ti_2AlX$ possesses a hexagonal nanolayered crystal structure with $P6_3$/mmc (#194) space group symmetry [11]. The unit cell of $Ti_2AlX$ contains a total of eight atoms, including four Ti, two Al, and two $X$ atoms, corresponding to two formula units. Wyckoff positions for Ti, Al, and $X$ (B, C, and N) atoms are $4f$ (1/3, 2/3, $z$), $2d$ (1/3, 2/3, 3/4), and $2a$ (0, 0, 0), respectively, where $z$ equals 0.087 [9]. Our simulated lattice parameters, volume, and density for all the studied compounds at ambient pressure are in very good agreement with previous experimental [5,6,39] and theoretical studies [9,11,40].

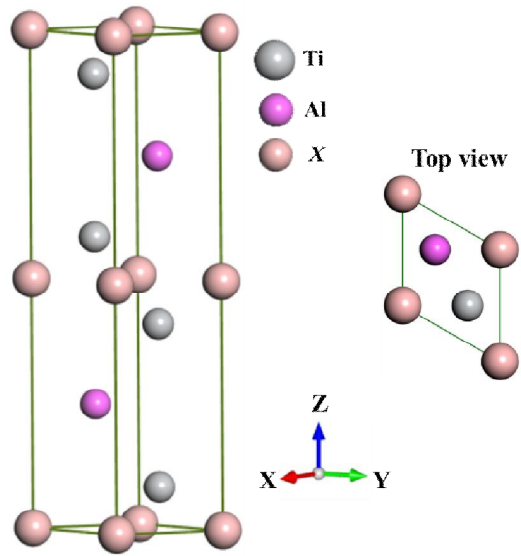

**Fig. 1.** The unit cell structure of $Ti_2AlX$ ($X$ = B, C, and N) MAX phases.

The pressure-dependent lattice parameters, their ratio (*c*/*a*), volume, and density data at 0 K up to 10 GPa with a 2 GPa step are collected and presented in Fig. 2. At ambient pressure, lattice constants and volumes decrease as we go from B to N (B→C→N) in accordance with the mass of the anions, while densities increase. This sequence is maintained under applied pressure. Therefore, as the pressure increases, the bond length shortens and the interaction between the elements (Ti, Al, and *X*) becomes stronger. The value of the *c*/*a* ratio is a significant indicator of whether our materials are kinking nonlinear elastic (KNE) solids or not [41]. The *c*/*a* ratio for all the studied compounds is greater than 4. Therefore, we can surmise that our materials are KNE. Ti$_2$AlC has been already confirmed as a KNE material through experimental work by B. Poon *et al.* [42]. The *c*/*a* ratio is much higher for Ti$_2$AlN than for Ti$_2$AlB and Ti$_2$AlC. All the compounds are more compressible along *c*-axis than *a*-axis (see Fig. 2a, b, and c) indicating a decrease in interlayer distance along *z*-axis. No evidence of pressure-induced phase transition was found for all the compounds, as confirmed by the monotonous changes in structural parameters under pressure (see Fig. 2).

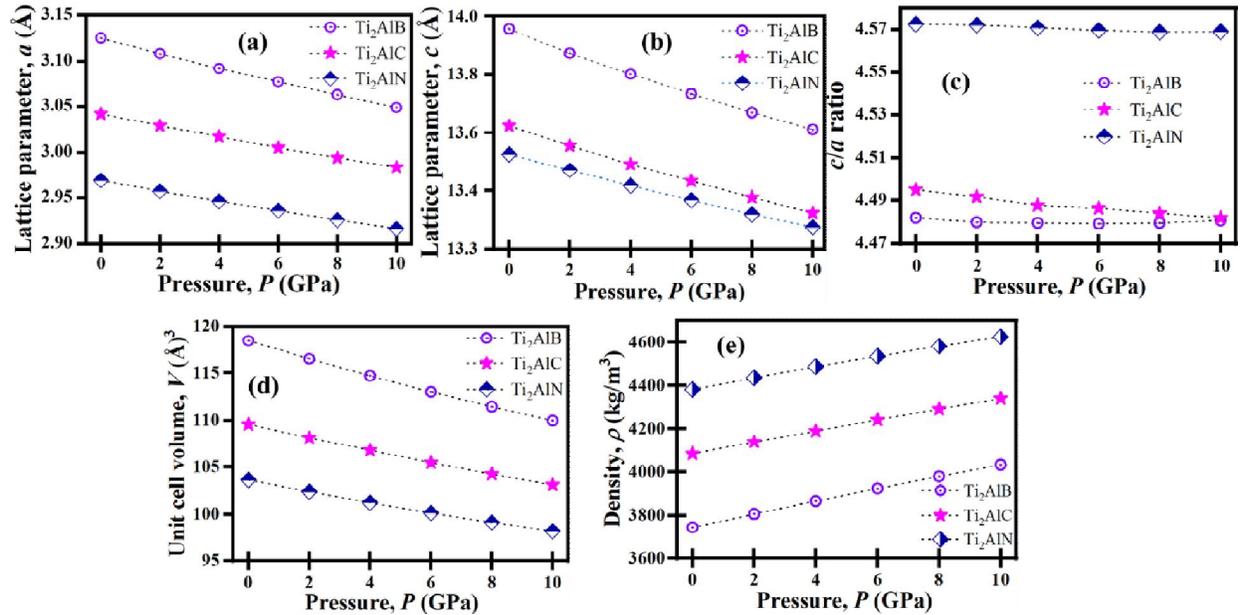

**Fig. 2.** The pressure-dependent lattice parameters (*a* and *b*), *c*/*a* ratio (a-c), volume (d), and density (e).

### 3.2. Mechanical and elastic properties

Information regarding the second-order elastic constants ($C_{ij}$) are important for exploring the mechanical behavior of materials and determining their potential for practical applications [43-45]. Elastic stiffness constants of solids change with both pressure and temperature. The obtained pressure dependent elastic constants are shown in Fig. 3(a, b, and c). The independent elastic constants for Ti$_2$Al*X* at ambient pressure align well with those reported in the literature [11,16]. All the compounds meet the Born stability criterion [46]: $C_{11} - |C_{12}| > 0$, $(C_{11} + C_{12})C_{33} - 2C_{13}^2 > 0$, $C_{44} > 0$, indicating that these MAX compounds are mechanically stable at ambient

pressure. Moreover, the mechanical stability criteria for hexagonal crystal under pressure (P) are as follows [46]:

$$\tilde{C}_{44} > 0, \tilde{C}_{11} > |\tilde{C}_{12}|, \tilde{C}_{33}(\tilde{C}_{11} + \tilde{C}_{12}) > 2\tilde{C}_{13}^2 \quad (2)$$

where, $\tilde{C}_{ii} = C_{ii} - P(i = 1, 3)$ and $\tilde{C}_{1i} = C_{1i} + P(i = 2, 3)$

All the elastic constants of the studied compounds satisfy the stability criteria under pressure. Therefore, all the compounds are mechanically stable up to 10 GPa.

The resistance to linear compression and atomic bonding strength along the [100] and [001] directions is directly related to $C_{11}$ and $C_{33}$, respectively [47]. Ti$_2$AlN exhibits greater resistance to both the linear compression and atomic bonding strength along the *a*- and *c*-axis, respectively than Ti$_2$AlB and Ti$_2$AlC. Unlike Ti$_2$AlB, the considerably larger values of $C_{11}$ than $C_{33}$ for both Ti$_2$AlC and Ti$_2$AlN indicate that the atomic bonds along [100] are stronger compared to that along [001] direction. On the other hand, the resistance to shape change is represented by $C_{12}$, $C_{13}$, and $C_{44}$. $C_{44}$ represents the resistance to shear deformation with respect to tangential stress applied to the (100) plane in the [010] direction of the compound [48]. The value of $C_{44}$ is also related to the valence electron density in the material; the higher the valence electron density, the higher the $C_{44}$. The stiffness order of the compounds under pressure is Ti$_2$AlN > Ti$_2$AlC > Ti$_2$AlB. Based on the definition of $C_{12}$ [44], Ti$_2$AlB exhibits greater shear resistance in the (100) plane along the *a*-axis than the other two materials. Furthermore, it is evident that $C_{ij}$ values for Ti$_2$Al*X* increase almost linearly with applied pressure. However, the rates of increase for $C_{11}$ and $C_{33}$ are faster than those for $C_{12}$, $C_{13}$, and $C_{44}$ (see Fig. 3a-c). Each of the elastic constant individually influences the properties of the materials, which we will discuss in the next sections.

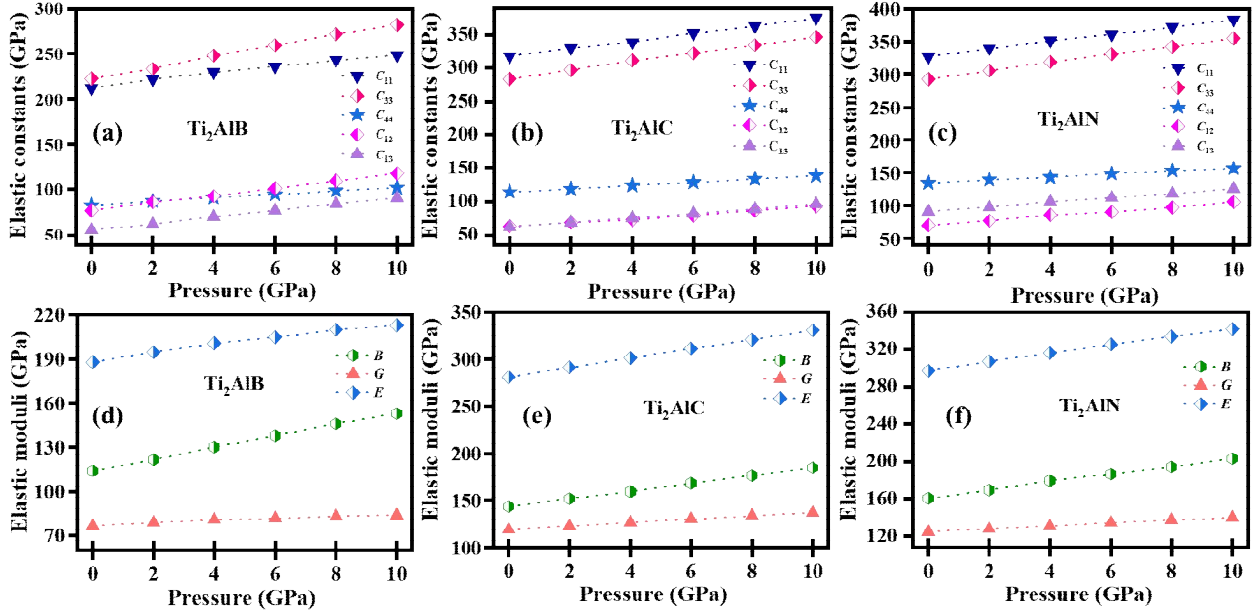

**Fig. 3.** Variation in the elastic constants and moduli with pressure for Ti$_2$AlB (a, d), Ti$_2$AlC (b, e), and Ti$_2$AlN (c, f) MAX phases.

Different elastic moduli are calculated from the $C_{ij}$ via the Voigt-Reuss-Hill (VRH) approximations [49-51]. The bulk ($B_V$, $B_R$, and $B_H$), shear ($G_V$, $G_R$, and $G_H$), and Young's ($E_V$, $E_R$, and $E_H$) moduli of the materials were estimated using the VRH formulae [47]. The Bulk modulus ($B$), shear modulus ($G$), and Young's modulus ($E$) of the Ti$_2$Al$X$ MAX phases from 0 to 10 GPa pressure range are displayed in Fig. 3d-f. The $B$ and $G$ of the material represent resistance to volume changes (bond length) and plastic deformation (bond angle) under external stress. The lower value of $G$ compared to $B$ for all studied materials suggests that plasticity will dominate their mechanical strength up to 10 GPa. The $E$ value of a material can quantify its stiffness and ability to resist thermal shock [52]. A larger $E$ value indicates greater stiffness, higher melting temperature, and lower resistance to thermal shock. Therefore, Ti$_2$AlN is much stiffer and has lower thermal shock resistance than Ti$_2$AlB and Ti$_2$AlC MAX phases, making them (Ti$_2$AlB and Ti$_2$AlC) more suitable as thermal barrier coating (TBC) system. All the elastic moduli for the compounds at 0 GPa are in good agreement with prior literature [11].

All the elastic moduli increase systematically with pressure and show no anomalies, corresponding to the phase stability of all three materials. Furthermore, a linear increase in moduli with pressure is observable in all the crystals, although their values vary, indicating the absence of any phase transitions. The pressure derivative bulk modulus ($B'$) for all three compounds is simulated and displayed in Fig. 4a. $B'$ increases as we move from B to N of Ti$_2$Al$X$ ($X$ = B, C, and N).

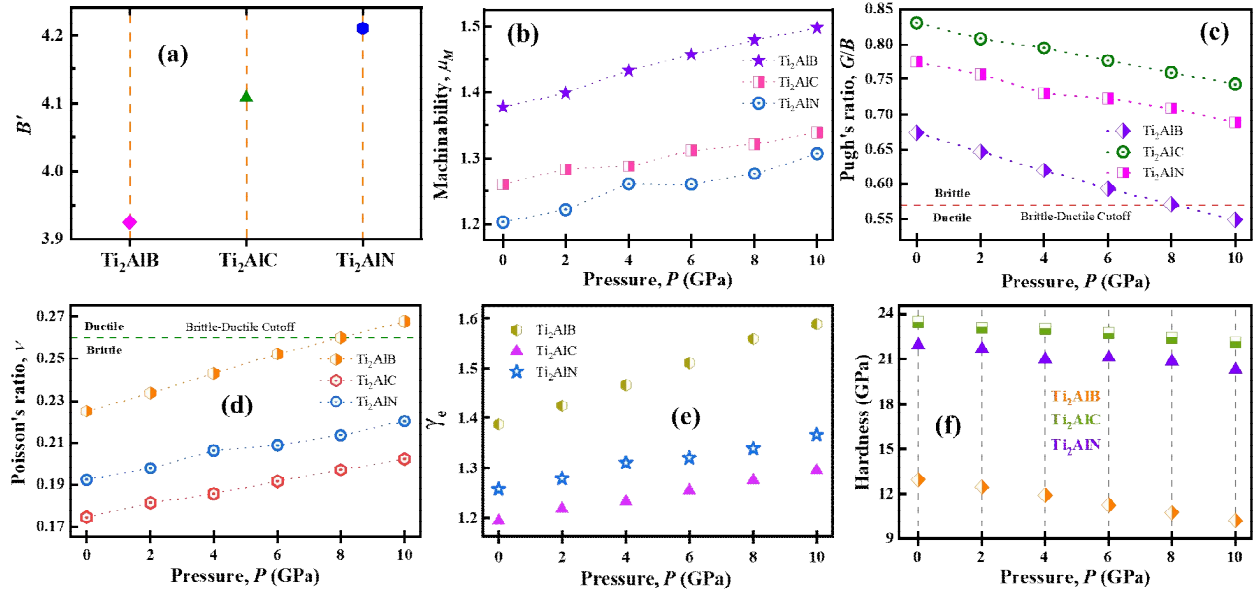

**Fig. 4.** The pressure derivate of bulk modulus (a), machinability (b), Pugh's ratio (c), Poisson's ratio (d), elastic Grüneisen parameter (e), and microhardness (f) of Ti$_2$Al$X$.

The assessment of the machinability index, a useful performance parameter of materials, has drawn significant interest in the development of manufacturing processes for complex cutting geometries of materials and tools [44]. The machinability index of a material is defined by $\mu_M = B/C_{44}$ [43]. The pressure-dependent $\mu_M$ of Ti$_2$Al$X$ is illustrated in Fig. 4b. At ambient pressure, the $\mu_M$ for Ti$_2$AlB is maximum and decreases as we move to a higher atomic number of $X$ (C→N), which is also aligned with $C_{44}$ value of the compounds. The machinability of the studied compounds increases with pressure. All the compounds possess a medium level of machinability comparable to many other well-known MAX phases [53].

Pugh stated that the ratio of $G$ and $B$ ($G/B$) is a measure of intrinsic ductility and brittleness. Pugh's ratio also addresses the interplay between brittle failure and shear failure within the material. Materials that deform through shear or tension under loading conditions are termed intrinsically ductile or brittle, respectively [54]. For brittle and ductile materials, the value of $G/B$ is higher and lesser than 0.57, respectively. Fig. 4c displays that both Ti$_2$AlC and Ti$_2$AlN are intrinsically brittle, failing under tensile stress, and they also retain their brittleness across a range of pressures considered here. On the contrary, Ti$_2$AlB exhibits intrinsic brittleness and shows a brittle-to-ductile transition for pressures above 6 GPa. Although a direct comparison between experimental and theoretical measures of ductility and brittleness is not straightforward, elongation and reduction of area critically depend on several factors, such as temperature, grain size, processing route (e.g., annealed vs. as-fabricated), impurity concentrations, and strain rate. The linear response of the studied materials under applied pressure indicates the absence of any kind of fracture.

Another well-known parameter that characterizes the mechanical behavior (compressibility, brittleness/ductility, and bonding nature/force) of materials is the Poisson's ratio ($\nu$). This parameter is estimated using the following expression: $\nu = (3B - 2G)/2(3B + G)$ [43]. The values of $\nu$ range between −1.0 and 0.5 (viscoelastic) for all elastically stable materials. Materials exhibit brittleness if $\nu$ is less than 0.26 and ductility if $\nu$ is greater than 0.26 [55]. Moreover, $\nu$ represents the lower (0.25) and upper (0.50) limits for central-force solids [47]. The borderline values of $\nu$ for covalent and ionic materials are 0.10 and 0.25, respectively [47]. The calculated Poisson's ratio suggests that all three MAX phase nanolaminates are brittle in the ground state with non-central inter-atomic forces and they are significantly covalently bonded at ambient pressure to up to 8 GPa (see Fig. 4d). With a gradual increase in $\nu$ as pressure rises, there is an implication of increased plasticity. Additionally, there is a brittle-to-ductile transition of $Ti_2AlB$ at approximately 8 GPa. The findings are also consistent with Pugh's ratio analysis. The lower $\nu$ of a material is associated with stronger and more directional covalent bonding and higher hardness (see Fig. 4d, and f) [56]. Therefore, $Ti_2AlC$ should have the strongest covalent bonding out of all three compounds with a sequence of $Ti_2AlC > Ti_2AlN > Ti_2AlB$.

The Grüneisen parameter for solids is a significant dimensionless thermodynamic parameter that has both pressure and temperature dependencies. This parameter can explain various physical phenomena such as thermal conductivity, thermal expansion, specific heat, elastic properties, and acoustic wave attenuation. Out of five Grüneisen parameters, we have estimated the elastic Grüneisen parameter ($\gamma_e$) for all three compounds using Poisson's ratio [45,57]: $\gamma_e = 3(1 + \nu)/2(2 − 3\nu)$. Fig. 4e shows the relationship between $\gamma_e$ and applied pressure. The value of $\gamma_e$ systematically increases with pressure. The order of $\gamma_e$ and $k_{ph}$ (phonon thermal conductivity) for $Ti_2AlX$ are in good agreement with the concept presented in an earlier work [47].

Potential materials for applications such as cutting tools, wear-resistant coatings, tribological coatings, solid lubricating coating, and heavy-duty products are chosen based on their hardness [44,58-60]. There are several macroscopic models for hardness calculation, and their efficiency depends on the crystal structure. These formulae have been summarized in our previous works [44,47]. The Vickers micro-hardness of hexagonal $Ti_2AlX$ compounds both at ambient and under pressure, is estimated from the elastic moduli using the formula [47,61]:

$$H = 2(k^2 G)^{0.585} − 3 \qquad (3)$$

here, $k \ (= G/B)$ is the Pugh's ratio and the moduli are in GPa. The estimated hardness of the compounds is displayed in Fig. 4f. The studied materials possess moderate hardness in the order: $Ti_2AlC > Ti_2AlN > Ti_2AlB$. The hardness of $Ti_2AlX$ decreases with increasing applied pressure, the decrease rate follows the hardness order, in contrast to their elastic moduli. The reported values here are rough predictions since there are many intrinsic and extrinsic factors that affect the hardness. Furthermore, variety of hardness measures exists giving different values. Solid lubricants are divided into two broad categories based on hardness: soft (H < 10 GPa) and hard (H >10 GPa) [60]. Solid lubricants are critically important for harsh operating conditions where liquid lubricants fail, such as at high vacuum, aerospace, high-speeds, high loads, and at very low or high temperatures.

Cauchy pressure ($P_C$) is another fundamental mechanical parameter that can explain brittleness/ductility and bonding nature. The $P_C$ for the studied materials (hexagonal) were estimated using the following expressions [62-64]:

$$P_C^a = (C_{13} - C_{44}) \text{ and } P_C^b = (C_{12} - C_{66}) \qquad (4)$$

where, $P_C^a$ and $P_C^b$ define the Cauchy pressure for the (100) and (001) planes, respectively. The estimated Cauchy pressures of the compounds are presented in Fig. 5(a, b). The value of $P_C$ is positive for metallic bonding. Besides, materials are expected to exhibit more directionality, bonding immobility, and brittleness if $P_C$ is negative [43]. The presence of metallic-like bonding along $c$-direction (positive $P_C^b$) in Ti$_2$AlB is an indicator of being more conducting than the other two (see Fig. 11d,e and 12a). On the other hand, $P_C^a$ is more negative for Ti$_2$AlB than Ti$_2$AlC and Ti$_2$AlN with the sequence: Ti$_2$AlC > Ti$_2$AlN > Ti$_2$AlB. Pettifor stated that the positive and negative $P_C$ of a material reflect metallic (non-directional) and angular bonding, respectively [62].

Studying mechanical anisotropy of materials is crucial for their application potentials in engineering sciences. Mechanical anisotropy controls several physical phenomena in solids (metal, glass, ceramics, etc.), such as micro-crack formation and its motion, the development of plastic deformation, and the relaxation of plasticity in thin-film materials. Therefore, we have studied the mechanical anisotropy of the MAX compounds under applied hydrostatic pressure. The universal anisotropy index ($A^U$), one of the most widely used anisotropy parameter, for the materials can be estimated from the following formula [65]:

$$A^U = 5\frac{G_V}{G_R} + \frac{B_V}{B_R} - 6 \geq 0 \qquad (5)$$

here, $B_V$, $G_V$, $B_R$, and $G_R$ represent the Voigt and Reuss approximated elastic moduli [47]. The Voigt (V) and Reuss (R) approximations define the upper and lower limits of $B$ and $G$ of a material, respectively. $A^U = 0$ for locally isotropic single crystals, whereas the degree of anisotropy extends with the deviation from zero. The $A^U$ of Ti$_2$Al$X$ as a function of applied pressure is depicted in Fig. 5c. At ambient pressure, all the studied compounds are anisotropic with the order: Ti$_2$AlB > Ti$_2$AlN > Ti$_2$AlC. The mechanical anisotropy of Ti$_2$AlN increases more rapidly with pressure than Ti$_2$AlB and Ti$_2$AlC. This implies that the chemical bonding properties are most sensitive to pressure in this compound.

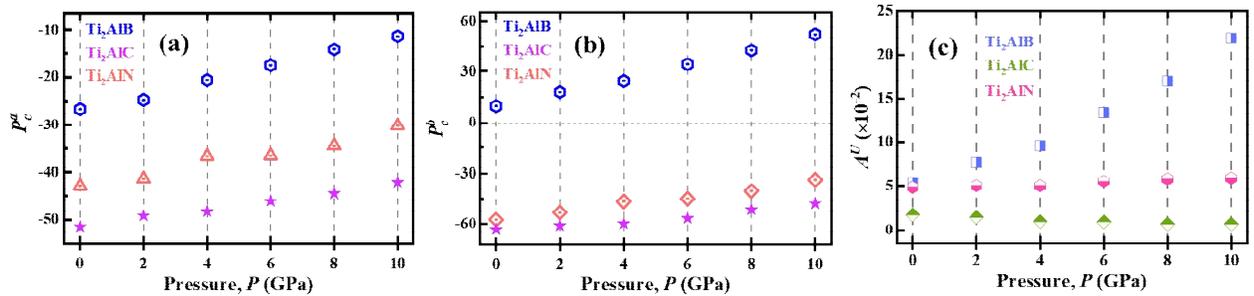

**Fig. 5.** The Cauchy pressure (a, b) and universal anisotropy index (c) of Ti$_2$Al$X$ ($X$ = B, C, and N) as a function of pressure.

To illustrate the anisotropy of Young's modulus ($E$), compressibility ($\beta$), Shear modulus ($G$), and Poisson's ratio ($\nu$) of the crystals at 0 GPa and under 10 GPa, 3D graphical visualization are also studied with the help of ELATE program [66]. The elastic stiffness matrixes, required as input, are estimated using the CASTEP code. Perfectly spherical shape is the manifestation of the isotropic nature of crystals. The 3D graphical representations of $E$, $\beta$, $G$, and $\nu$ for the compounds are displayed in Fig. 6-7. The maximum and minimum points of the parameters ($G$ and $\nu$) are represented by solid transparent blue and green colors, respectively. Ti$_2$AlC exhibits the least anisotropy in $\nu$ at 0 GPa. The Poisson's ratio anisotropy of Ti$_2$AlB increases more than Ti$_2$AlC and Ti$_2$AlN under 10 GPa which is also in agreement with the $A^u$ study (see Fig. 5c). All the four parameters exhibit isotropy along [100] direction at both 0 and 10 GPa. We can conclude that the materials possess anisotropy in the following order: $\beta < E < G < \nu$.

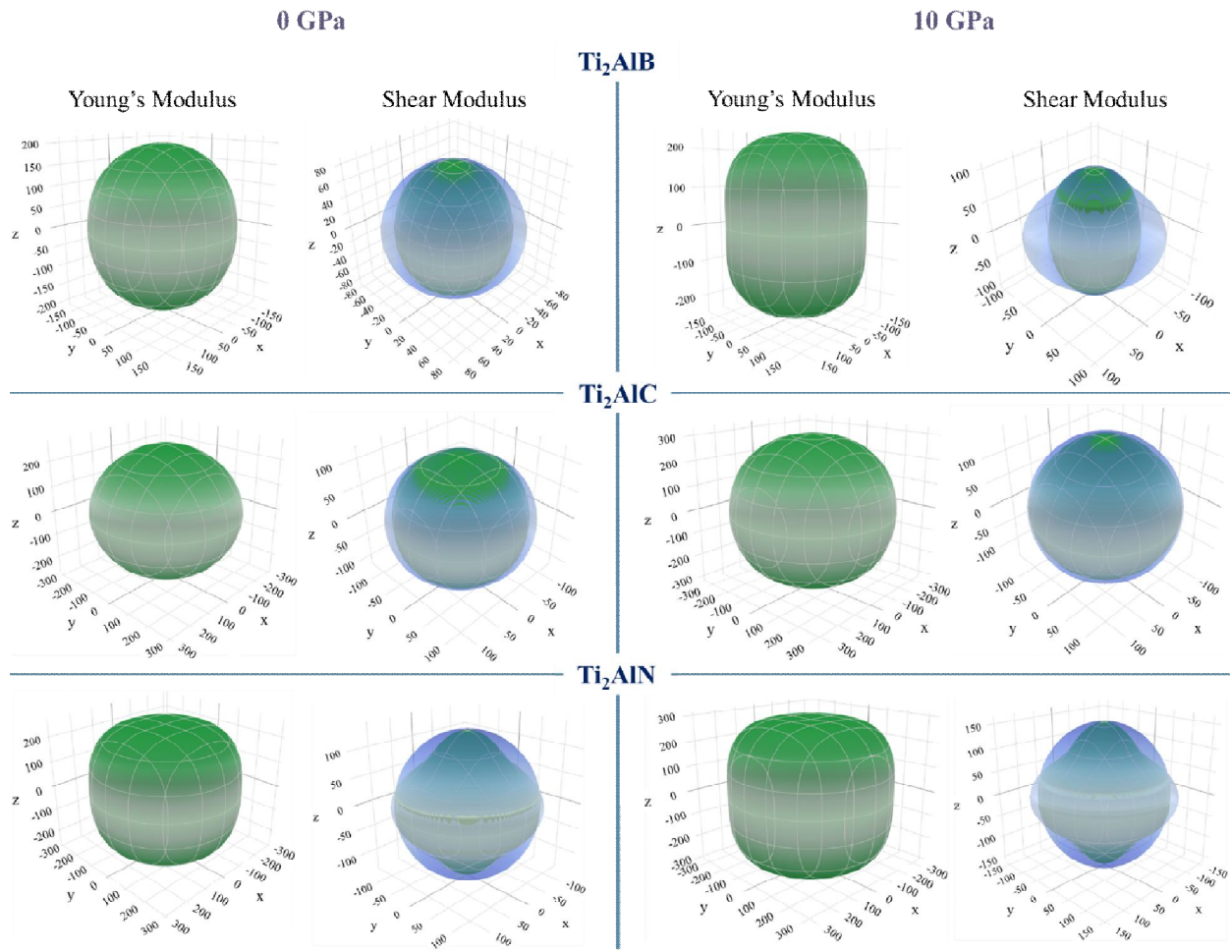

**Fig. 6.** Directional dependence of Young's modulus ($E$), and shear modulus ($G$) of Ti$_2$Al$X$ ($X$ = B, C, and N) at 0 and 10 GPa.

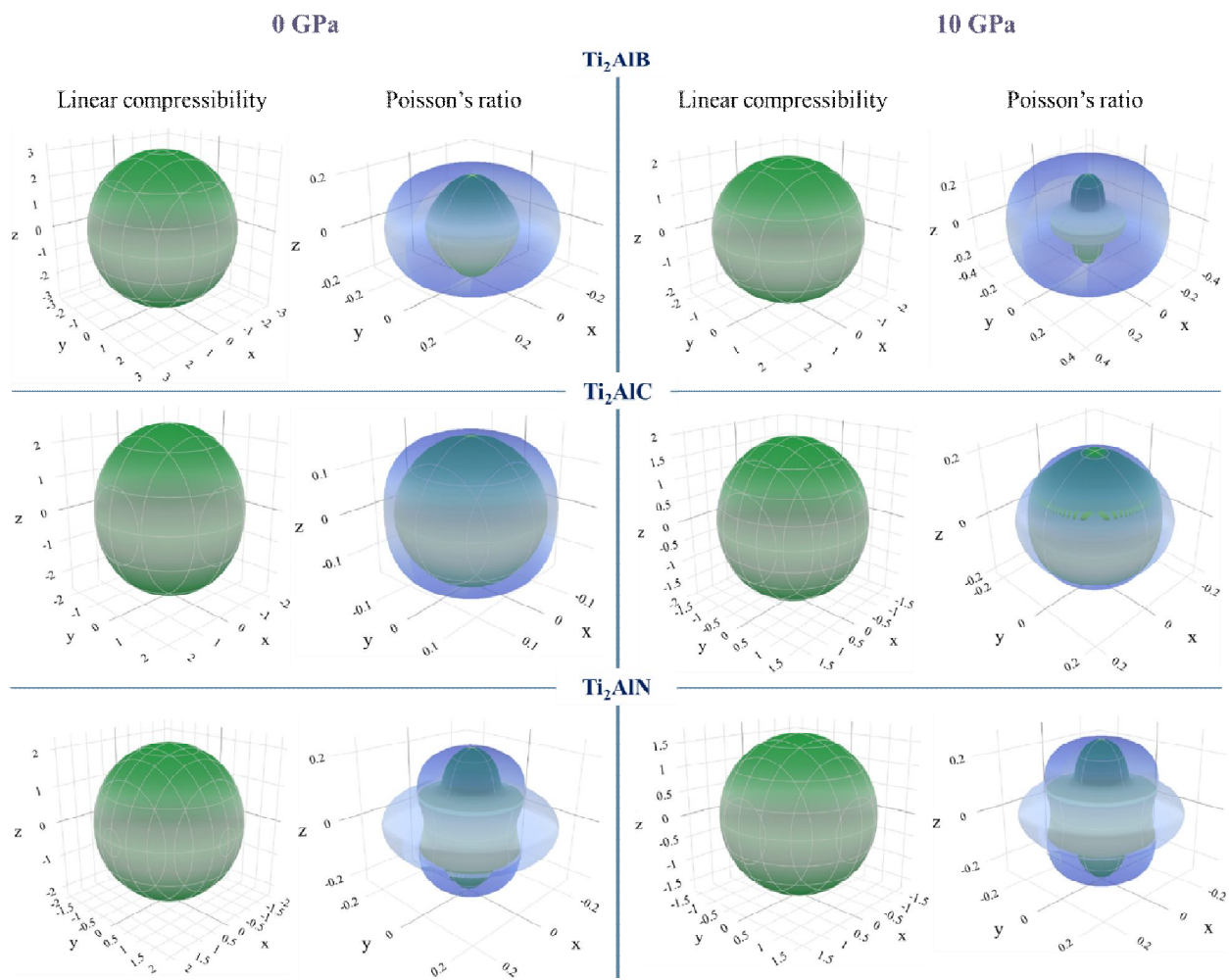

**Fig. 7.** Directional dependence of compressibility ($\beta$), and Poisson's ratio ($\nu$) of Ti$_2$Al$X$ ($X$ = B, C, and N) at 0 and 10 GPa.

## 3.3. Electron density difference

The interatomic bonding nature in a material can be explored from the electron density difference (EDD) around the different elements in the material, which displays the accumulation and depletion of charges around the elements. We have calculated EDD for Ti$_2$Al$X$ at 0 GPa and 10 GPa, as displayed in Fig. 8. The color scale for each compound's map illustrates the electron density difference. For Ti$_2$AlB and Ti$_2$AlC, the red and green color indicates high and low charge (electron) density, respectively, whereas vice versa for Ti$_2$AlN. Maximum EDD around $X$ atoms for all the studied compounds illustrates that electrons/charges are accepted by this element. On the other hand, the blue color (red for Ti$_2$AlN) Ti and Al atoms signify electrons are donated by them. The covalent bonding between Ti-$X$ and Al-$X$ atoms is confirmed. Out of the three compounds, Ti$_2$AlC possesses the least EDD value. There is a clear directional dependency of EDD for all the studied compounds.

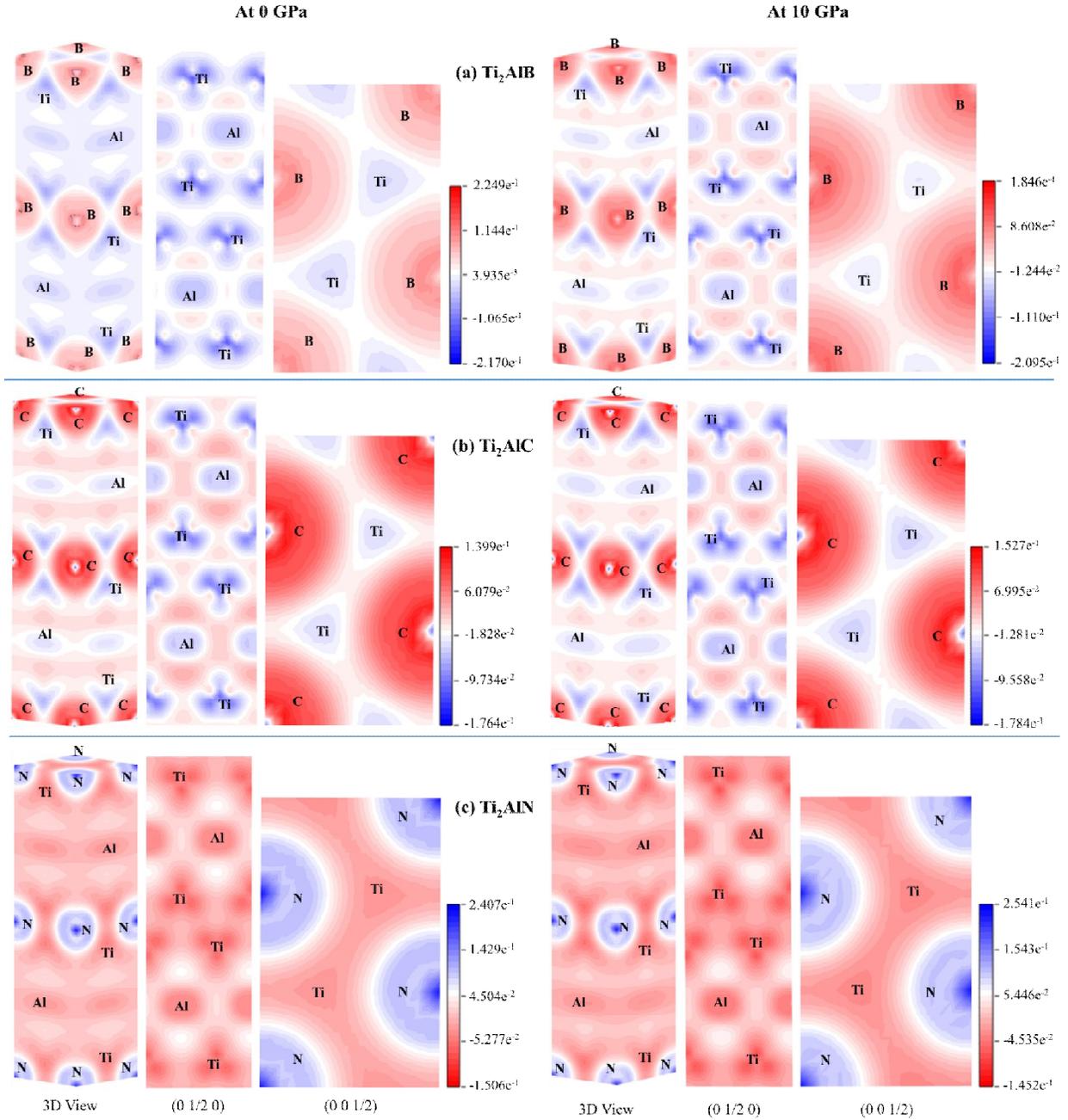

**Fig. 8.** The electron density difference (Å$^{-3}$) maps for (a) Ti$_2$AlB, (b) Ti$_2$AlC, and (c) Ti$_2$AlN in different planes under 0 GPa and 10 GPa hydrostatic pressures.

### 3.4. Acoustic properties

The transverse (shear), longitudinal, and average sound velocities through the studied materials have been estimated using well-known formulae from the literature [67,68]. The hydrostatic pressure dependence of the velocities of the acoustic waves propagating through the compounds is also studied, as displayed in Fig. 9. In general, the longitudinal sound velocity is greater than the transverse one. Lower values of $v_t$ compared to $v_l$ in all the three compounds indicate that the near-surface information will be dominated by transverse wave. Acoustic velocity through a material is directly related to its density, uniaxial compression strength, Young's modulus, and thermal conduction ability. The longitudinal sound wave velocity increases more rapidly with pressure than the other two wave velocities for all the studied compounds which also correlates with the other parameters' response with pressure.

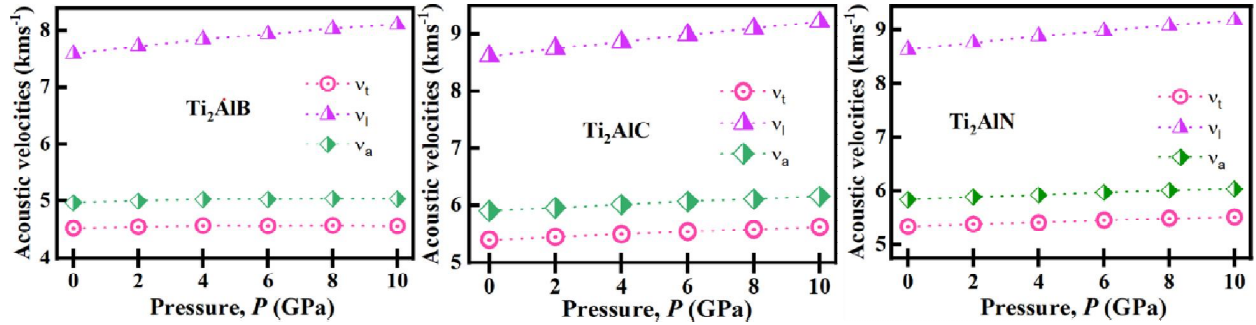

**Fig. 9.** The transverse ($v_t$), longitudinal ($v_l$), and average ($v_a$) sound velocities for Ti$_2$Al$X$ ($X$ = B, C, and N) compounds under applied pressure.

### 3.5. Phonon and vibrational analyses

Both the electrons and phonons in a material conduct heat. In solids, acoustic phonon and optical phonon are not independent but interact with each other like, acoustic-acoustic phonon, acoustic-optical phonon, and optical-optical phonon. Fig. 10 shows the calculated phonon dispersion with their corresponding total and partial density of states (in the right panel of each diagram) in the BZ of Ti$_2$AlB, Ti$_2$AlC, and Ti$_2$AlN, at ambient and under 10 GPa pressure. The absence of any negative energy phonon branch for the studied compounds ensures their dynamical stability both at 0 and 10 GPa. In fact, we have found dynamical stability at all the pressures up to 10 GPa. A material with N atoms per unit cell possesses 3 acoustic modes and (3N-3) optical modes. The eight atoms in the unit cell of Ti$_2$Al$X$ indicates total of 24 vibrational modes, with 3 acoustic and 21 optical branches: $6E_{1u} + 3A_{2u} + 4E_{2u} + 2B_{1u} + 4E_{2g} + 2B_{2g} + 2E_{1g} + A_{1g}$. The low-frequency region consists of one longitudinal mode (in-plane) and two transverse acoustic modes (one in-plane and one out-of-plane). Acoustic phonon modes dominate the transport of sound and thermal energy in crystals [44]. Coherent vibrations of atoms in a lattice outside their equilibrium position produce acoustic branches. The acoustic vibrational modes of Ti$_2$Al$X$ correspond to the $2E_{1u}$ and $A_{2u}$ modes. Moreover, optical vibrational modes can both be transverse and longitudinal. Out of the 21 optical modes, 6 are infrared (IR)-active and 7 are Raman-active, while the rest are inactive to IR and Raman. These active modes are: $4E_{1u} + 2A_{2u} + 4E_{2g} + 2E_{1g} + A_{1g}$. The different phonon modes, both active and silent modes, of all the three compounds at ambient pressure are listed in Table 1. These mode frequencies will be useful for the experimental confirmation of the phases later on, especially for Ti$_2$AlB. The different phonon frequencies of optical branch around $\Gamma$-point indicate the ionic bonding [22]. The phononic bandgap defines the frequency region where elastic waves through the material are inhibited. The splitting of optical branches (lower and upper) is due to the mass differences among atoms. The overlap/hybridization between acoustic and lower optical branches for all the compounds indicates absence of any phononic bandgap (elastic impedance) between these two branches. Dispersion of phonon modes is related with the bond lengths between atoms in the materials.

Sharp peaks in total phonon density of states (PHDOS) indicate flat phonon dispersion. The total PHDOS in the higher optical branch displays sharp peaks as we move from B to N element. The broad peak in the PHDOS of $Ti_2AlB$ indicates non-localized phonon dispersion compared to the other two compounds. Partial PHDOS for all the compounds confirms that the acoustic and lower optical modes are dominated by vibration of both Ti and Al atoms. Contribution of Ti element is more significant than Al. On the other hand, high frequency optical modes (~16 to ~22 THz) are mainly generated from the lightest elements such as B, C, and N in the MAX compounds considered. The vibrational frequency of the elements in a material (assuming that the oscillations are harmonic) can be expressed as: $\omega = \sqrt{k/m}$; where, $k$ is the bond stiffness and $m$ is the mass of the element. The strong Ti-$X$ bonding makes the frequency of $X$ atoms much higher than that of the Ti and Al atoms. The heat capacity of a material is proportional to the PHDOS [69].

The phonon frequency for all the compounds increases with pressure. This is because the interaction between atoms becomes stronger as the interatomic distance decreases under pressure, leading to an increase in phonon frequency.

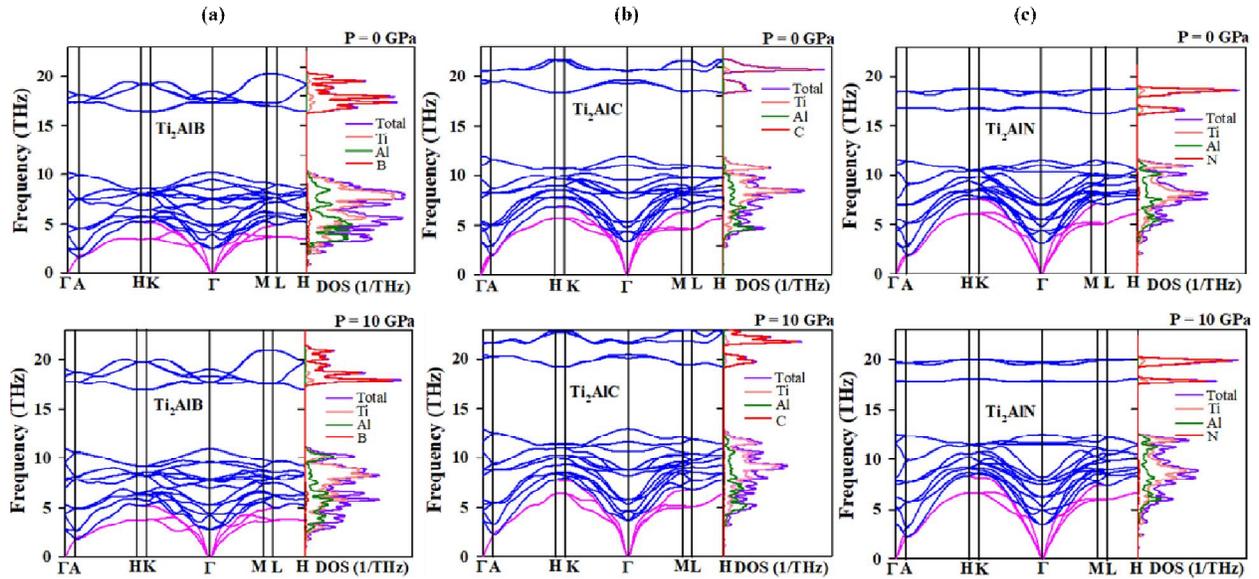

**Fig. 10.** The phonon dispersion spectra (PDS) and phonon density of states (PHDOS) of (a) $Ti_2AlB$, (b) $Ti_2AlC$, and (c) $Ti_2AlN$ under 0 GPa and 10 GPa hydrostatic pressure.

**Table 1:** Calculated values of Raman-active and infrared (IR) optical modes at the $\Gamma$-point of Ti$_2$AlX (X = B, C, and N) at 0 GPa.

| Theoretical Mode symmetry | Activity (cm$^{-1}$) | | | | | |
|---|---|---|---|---|---|---|
| | Raman | | | IR | | |
| | Ti$_2$AlB | Ti$_2$AlC | Ti$_2$AlN | Ti$_2$AlB | Ti$_2$AlC | Ti$_2$AlN |
| E$_{1u}$ | – | – | – | – | – | – |
| E$_{1u}$ | – | – | – | – | – | – |
| A$_{2u}$ | – | – | – | – | – | – |
| E$_{2u}$ | – | – | – | – | – | – |
| E$_{2u}$ | – | – | – | – | – | – |
| B$_{1u}$ | – | – | – | – | – | – |
| E$_{2g}$ | 139.190 | 147.112 | 147.997 | – | – | – |
| E$_{2g}$ | 139.190 | 147.112 | 147.997 | – | – | – |
| E$_{1u}$ | – | – | – | 163.086 | 177.797 | 185.265 |
| E$_{1u}$ | – | – | – | 163.086 | 177.797 | 185.265 |
| B$_{2g}$ | – | – | – | – | – | – |
| E$_{1g}$ | 249.225 | 276.293 | 231.403 | – | – | – |
| E$_{1g}$ | 249.225 | 276.293 | 231.403 | – | – | – |
| E$_{2g}$ | 252.453 | 147.112 | 235.393 | – | – | – |
| E$_{2g}$ | 252.453 | 147.112 | 235.393 | – | – | – |
| A$_{2u}$ | – | – | – | 282.050 | 317.871 | 344.145 |
| A$_{1g}$ | 315.822 | 367.656 | 366.615 | – | – | – |
| B$_{2g}$ | – | – | – | – | – | – |
| E$_{2u}$ | – | – | – | – | – | – |
| E$_{2u}$ | – | – | – | – | – | – |
| E$_{1u}$ | – | – | – | 578.355 | 684.892 | 614.821 |
| E$_{1u}$ | – | – | – | 578.355 | 684.892 | 614.821 |
| A$_{2u}$ | – | – | – | 592.401 | 640.847 | 559.393 |
| B$_{1u}$ | – | – | – | – | – | – |

### 3.6. Thermo-physical parameters

The thermal behavior of MAX phases was studied from Debye temperature, melting temperature, lattice thermal conductivity, minimum thermal conductivity, and thermal expansion coefficient. All the thermal parameters of Ti$_2$AlX (X = C, B, and N) studied here have been estimated from well-established formulae summarized elsewhere [43,47,67,70-72].

#### 3.6.1. Debye temperature

Studying Debye temperature ($\Theta_D$) of a solid is important in the material selection process for various applications related to lattice vibrations. The $\Theta_D$ value reflects many key properties of a solid, such as the bonding strength between elements, vibrational energy (thermal, acoustic, optical), melting temperature, indentation hardness, isothermal compressibility, specific heat capacity, superconducting transition temperature, electron-phonon coupling constant, etc [47]. There are two types of lattice waves: acoustic and optical. The Debye temperatures of Ti$_2$Al$X$ ($X$ = B, C, and N) MAX phases have been estimated using Anderson model given in the literature [67,68]. The Debye temperature versus applied pressure for all three compounds is displayed in Fig. 4e. A linear increase in $\Theta_D$ with applied pressures is observable for all the studied compounds, which is an indicator of thermodynamic stability of structures up to 10 GPa. The $\Theta_D$ of Ti$_2$AlB is much lower than the other two compounds, which makes it more suitable as thermal barrier coating (TBC). The pressure response of $\Theta_D$ for the studied compounds follows the trend of elastic moduli, melting temperature, minimum thermal conductivity, and acoustic impedance.

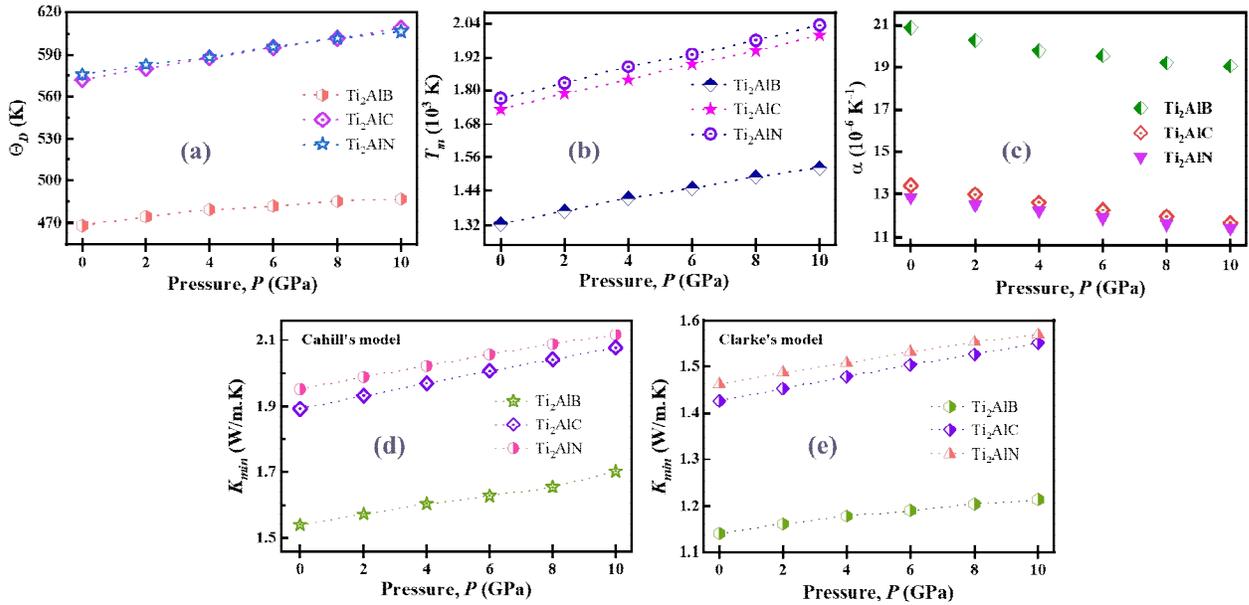

**Fig. 11.** (a) Debye temperature, (b) melting temperature, (c) thermal expansion coefficient, and (d,e) minimum thermal conductivity of the materials as a function of hydrostatic pressure.

### 3.6.2. Melting temperature

Melting temperature ($T_m$) of solids is crucial for exploring potential materials for various industrial applications [47]. Fig. 11b represents the effects of pressure on $T_m$ for Ti$_2$AlB, Ti$_2$AlC, and Ti$_2$AlN. It is found that $T_m$ for all three compounds are above 1000 ºC and increases monotonically with pressure, which indicates their good thermal stability. Ti$_2$AlC and Ti$_2$AlN possess higher $T_m$ value as compared to Ti$_2$AlB which is due to their relatively higher elastic constants and stronger atomic bondings. In addition, a high melting point arises from a high heat of fusion, a low entropy of fusion, or from a combination of both.

### 3.6.3. Thermal expansion coefficient

Materials go through dimensional (shape, area, and volume) changes under pressure, temperature, magnetic field, etc. Therefore, this has critical implications for materials used in industrial applications. Thermal expansion coefficient (TEC) for a material can quantify many phenomena, such as thermal conductivity, specific heat, entropy, and isothermal compressibility. Studying the TEC of materials is of interest for identifying potential materials for various applications in materials science including automotive, aerospace, railway lines/tracks, communication cables, anti-thermal shock system, electronic packaging, and spintronics [47].

The pressure effect on the TEC of the materials is studied here and displayed in Fig. 11c. The order of α for the materials at ambient pressure is: Ti$_2$AlB > Ti$_2$AlC > Ti$_2$AlN, with Ti$_2$AlB exhibiting a much higher TEC than the other two materials. A linear decrease in TEC with increasing pressure has been observed for all three compounds.

### 3.6.4. Minimum thermal conductivity

The minimum thermal conductivity ($k_{min}$) of materials describe the lowest limit to lattice thermal conductivity at temperature above the Debye temperature. Both the temperature and pressure can manipulate $k_{min}$ of a solid. The isotropic $k_{min}$ of the studied materials is stimulated from two well-known models [73,74]:

$$\text{Cahill's model: } k_{min} = \frac{k_B}{2.48} n^{2/3}(v_l + 2v_t) \quad (6)$$

$$\text{Clarke model: } k_{min} = k_B v_a (V_{atomic})^{-2/3} \quad (7)$$

here, $n$ is the number of atoms per unit volume and $V_{atomic}$ is the cell volume per atom. The estimated values of $k_{min}$ as a function of pressure are displayed in Fig. 11(d,e). Both the models predict that Ti$_2$AlB exhibits the lowest $k_{min}$ among all three and their values increase with pressure. Similar trend is also found in $\Theta_D$ and $T_m$. Moreover, Cahill's model predicts larger $k_{min}$ compared to Clarke model which is also observed for other materials [43,47,48].

### 3.6.5. Phonon thermal conductivity

Thermal conductivity in solids is the combination of electron and phonon thermal conductivity ($k_{ph}$). Both high and low phonon thermal conducting materials have industrial importance. Ultrahigh $k_{ph}$ materials are reported to exhibit large phonon group velocity, long phonon lifetime and strong bonding [75]. Materials with lower $k_{ph}$ show better thermoelectric behavior. For high temperature applications, a material should have a low thermal conductivity at higher operating temperature. Moreover, the decrease in intrinsic/extrinsic resistance to phonon-phonon and electron-phonon interactions in solids indicates decrease in $k_{ph}$. The lower value of $k_{ph}$ indicates the presence of softer phonon modes. The $k_{ph}$ of Ti$_2$Al$X$ ($X$ = C, B, and N) as function of temperature and pressure is estimated from well-known formulae reported in the literature [44,71,72] and displayed in Fig. 12.

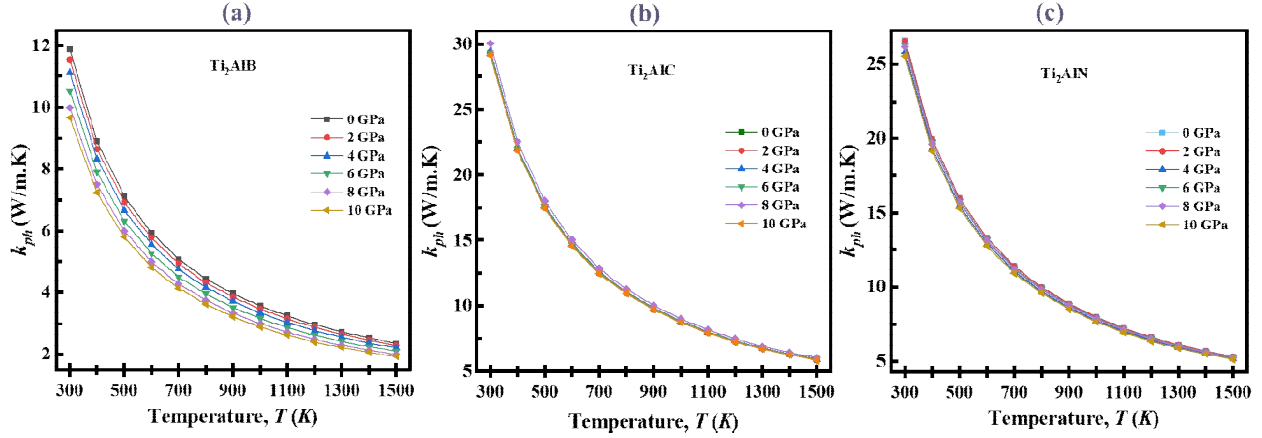

**Fig. 12.** The pressure and temperature dependence of $k_{ph}$ of (a) Ti$_2$AlB, (b) Ti$_2$AlC, and (c) Ti$_2$AlN MAX phases.

The $k_{ph}$ value of the compounds at 0 GPa and 300 K follows the sequence of Ti$_2$AlC > Ti$_2$AlN > Ti$_2$AlB. The value of $k_{ph}$ decreases gradually with increasing pressure and temperature for all three compounds. At ambient pressure and room temperature, Ti$_2$AlC has higher $k_{ph}$ than Ti$_2$AlB and Ti$_2$AlN which make it better thermal conductor. The $k_{ph}$ of Ti$_2$AlC shows the least response to pressure and temperature due to the elevated hardness found in this compound as discussed in mechanical and elastic properties section (see section 3.2). The pressure dependence of $k_{ph}$ follows this sequence: Ti$_2$AlB < Ti$_2$AlC < Ti$_2$AlN. Additionally, Ti$_2$AlB exhibits stronger pressure dependence than Ti$_2$AlC and Ti$_2$AlN, which also correlates with the $C_{ij}$, $E$, $\Theta_D$, and $T_m$ of the corresponding compounds. The $k_{ph}$ value of all the compounds decreases with temperature in the following order: Ti$_2$AlB < Ti$_2$AlN < Ti$_2$AlC, which correlates with acoustic and dynamic behavior studies. Ti$_2$AlB shows the lowest $k_{ph}$ and a weaker temperature response than the other two compounds, which may indicate a better heat resistance.

## 3.7. Electronic properties
### 3.7.1. Band structure and density of states

The calculated pressure-dependent electronic band structure of the studied compounds along the high-symmetry directions in the Brillouin zone is displayed in Fig. 13. The horizontal dashed line designates the Fermi level ($E_F$). The overlap of valence and conduction bands at $E_F$ for Ti$_2$AlB, Ti$_2$AlC, and Ti$_2$AlN indicates their metallic character. The states close to the Fermi level control the charge transport properties of materials. The anisotropy in electric conductivity of the studied compounds is obvious form the band energy dispersion around the $E_F$. Highly dispersive bands close to the Fermi level is indicative of high mobility of the charge carriers [76]. The most localized bands are along A-H path, whereas the highest band dispersion is observable along K-$\Gamma$-M path for all the compounds. The bands located along the A-H path indicate that they are less mobile or more tightly bound in that direction, resulting in restricted electron mobility. On the other hand, the high dispersion along the K-$\Gamma$-M path suggests that the electrons have more freedom and may perform better electrically in that direction. The $E_F$ of Ti$_2$Al$X$ lies below the valence band maximum near $\Gamma$-point. The substitution of $X$ with higher atomic number (B→C→N) in Ti$_2$Al$X$, introduces extra valence electrons per atom, leads $E_F$ moves to a higher energy level. Therefore, we can say that increase in bonding states near the $E_F$. Strongly hybridized bands consisting of Ti 3$d$ and Al 3$p$ orbitals spread over an energy range from − 2 eV to 2 eV. These bands contribute significantly in formation of chemical bonds and optical properties in the visible region. Bands associated with N are lower in energy than B and C, which is associated with their electrostatic attraction. Consequently, the mechanical strength of the materials also increases with $X$ (see section 3.2, mechanical behavior). Additionally, all three compounds remain metallic under pressure and the influence of pressure is not pronounced. We have also explored the effects of spin-orbit coupling on the electronic band structure. The overall effect was found insignificant (not shown here).

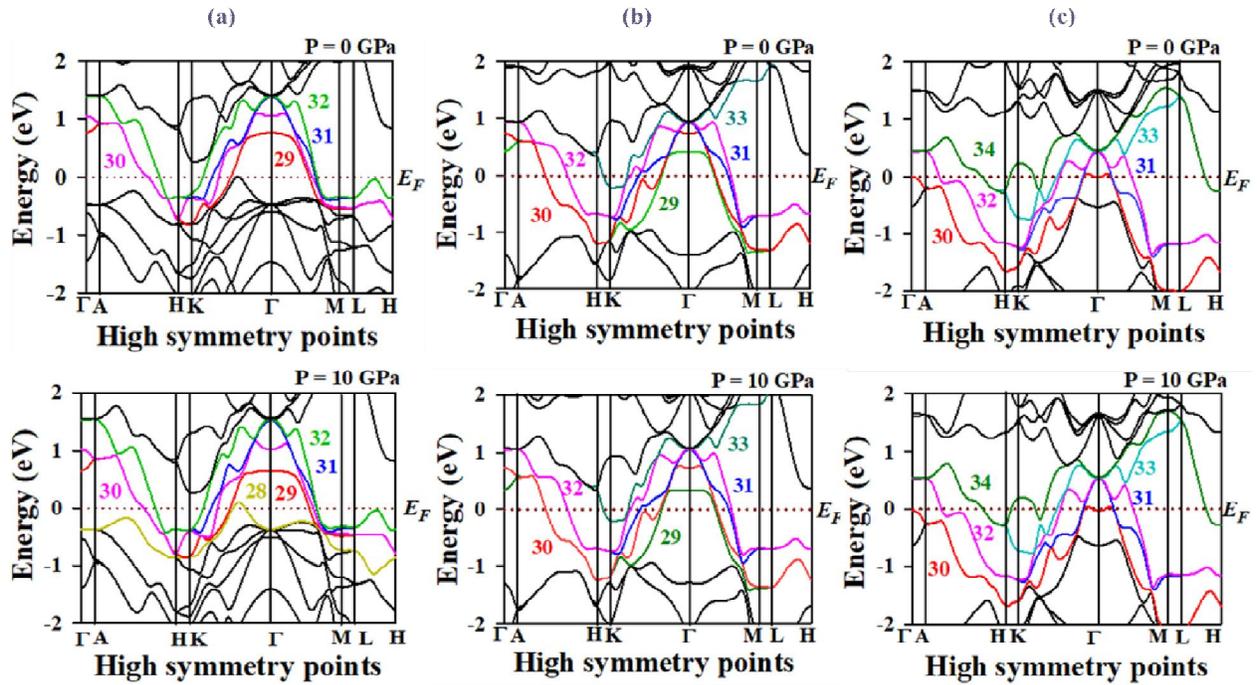

**Fig. 13.** The electronic band structure of (a) Ti$_2$AlB, (b) Ti$_2$AlC, and (c) Ti$_2$AlN along high symmetry directions of the Brillouin zone under 0 GPa and 10 GPa.

To gain further insights into the electronic responses of the compounds at P = 0 GPa and under hydrostatic pressure, we have studied the electronic density of states of the three MAX phases. The calculated pressure-dependent total density of states (TDOS) as well as partial density of states (PDOS) for Ti$_2$Al$X$ ($X$ = B, C, and N) are displayed in Fig. 14. The vertical dotted line represents Fermi energy ($E_F$). The strong hybridization between Ti 3$d$ and Al 3$p$ states is evident for all the compounds. The contribution of $X$ atoms to TDOS in Ti$_2$Al$X$ around the $E_F$ decreases with increasing atomic number. This might be understood this way, compared to lighter elements, heavier elements may have more inner-shell electrons, which are less engaged in bonding and electronic states close to the Fermi level. It is also important to point out that the TDOS below the $E_F$ is maximum for Ti$_2$AlB due to the lower band energy dispersion $E(k)$ in this energy range (see Fig. 13). Around $E_F$, the contribution of Ti 3$d$ orbitals mainly dominates in all the compounds, whereas that of the $X$ elements decreases as we move to heavier elements.

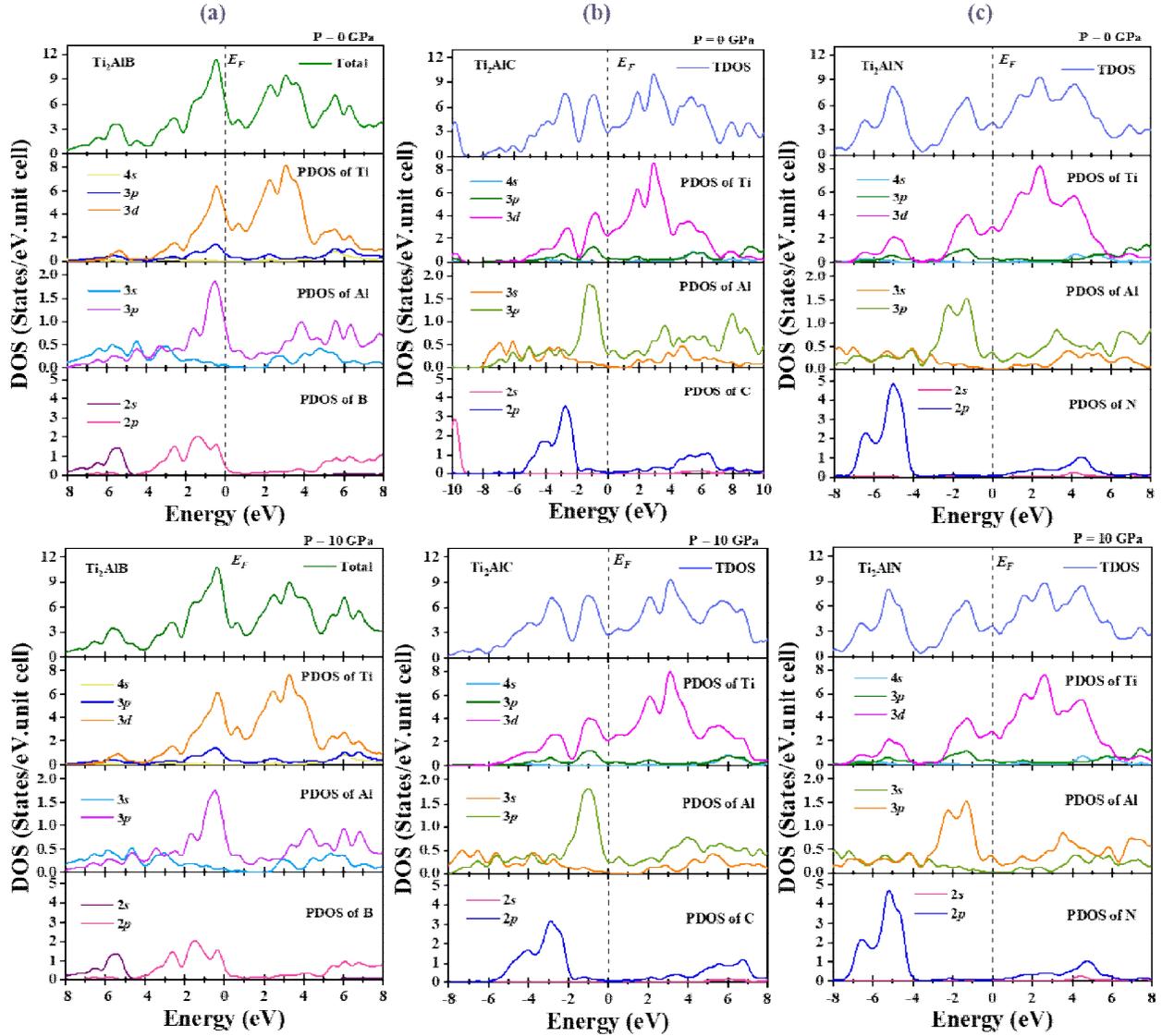

**Fig. 14.** Total and partial electronic density of states (TDOS and PDOS) of (a) Ti$_2$AlB, (b) Ti$_2$AlC, and (c) Ti$_2$AlN under 0 GPa and 10 GPa.

The TDOS value for all the three compounds at $E_F$ under pressure is presented in Fig. 15a. The TDOS at $E_F$ at ambient pressure follow the order of Ti$_2$AlB > Ti$_2$AlN > Ti$_2$AlC which is in good agreement with literature [9,20]. Therefore, Ti$_2$AlB is more metallic and the metallicity increases with pressure. On the contrary, metallicity of Ti$_2$AlC and Ti$_2$AlN decreases with increasing pressure. In superconductors, transition temperature is connected with their electronic density of states at $E_F$ [77].

The Coulomb pseudopotential ($\mu^*$), which is electron–electron interaction due to Coulomb force, plays an important role in determining electronic correlations in a system [78]. We have gauged $\mu^*$ using the following expression [79]:

$$\mu^* = \frac{0.26 N(E_F)}{1 + N(E_F)} \tag{8}$$

The variation in the calculated Coulomb pseudopotential for all three compounds with pressure is displayed Fig. 15b. The values of $\mu^*$ high in the ground state and vary weakly with pressure. The strongest electronic correlations exist in Ti$_2$AlB.

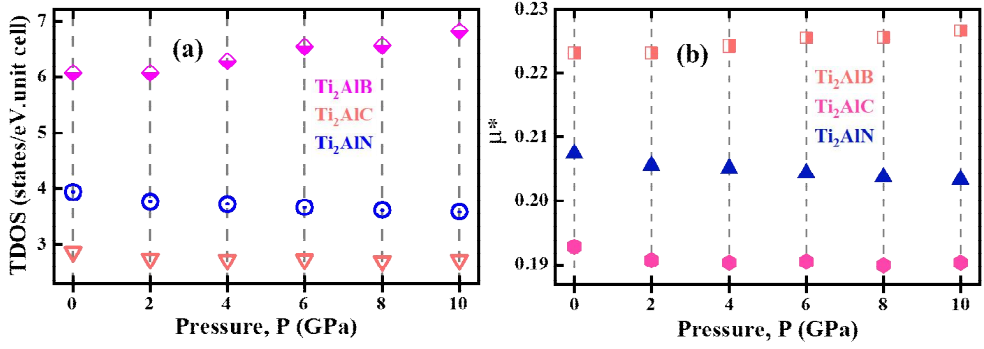

**Fig. 15.** The (a) total density of states and (b) Coulomb pseudopotential of Ti$_2$Al$X$ under pressure.

### 3.7.2. Fermi surface topology

The Fermi surface (FS) is crucial as it can explain several behaviors of a solid such as the electrical, magnetic, thermal, and optical properties. The dynamical properties of an electron on the FS largely depend on the position and shape of the Fermi sheets in the Brillouin zone. The effective mass of the electron is closely correlated with the FS curvature. The effective mass of the electron is smaller in areas where the FS is significantly curved, which can impact the electron's mobility and response to external fields. On the other hand, higher effective masses are associated with flat regions of the FS. The Fermi velocity may be relatively constant if the FS is approximately spherical; but, if it is strongly anisotropic, the velocity may vary dramatically in various directions. Nesting features, or regions on the FS where distinct parts can be superimposed by a certain wave-vector, promotes spin density waves or charge density waves. Anisotropic electrical and thermal conductivities can result from direction-dependent electron mobility in materials whose FS is extended along specific directions [80,81]. Electrons near the Fermi sheets are actively involved in the superconducting state formation [82]. The electrical characteristics of metallic systems may be inferred from the FS contour with respect to the BZ. Moreover, the dynamical characteristics of an electron on the FS are often location-dependent. The presence of an FS is probably the most important indications of materials containing Fermi liquid quasiparticles. The Fermi surfaces of Ti$_2$AlB, Ti$_2$AlC, and Ti$_2$AlN compounds were constructed from the bands crossing the $E_F$ at pressure of 0 and 10 GPa and are depicted in Figs. 16-18. The FS of Ti$_2$AlB has four sheets, while the other two materials contain five sheets.

For Ti$_2$AlB, all four sheets are central and parallel to the $\Gamma$–$A$ direction with cylindrical or prismatic-like hexagonal cross section at 0 GPa. These sheets are electron-like and 2D. Under 10 GPa, electron-like spherical FS (for band 28) appears along $\Gamma$–$K$ path, whereas other Fermi sheets remain identical.

For Ti$_2$AlC, four cylindrical electron-like Fermi sheets are constructed from band numbers 29, 30, 31, and 32 keeping the $\Gamma$–$A$ direction as their axis. All these sheets exhibit hexagonal cross section as Ti$_2$AlB. For band 33, electron pockets appear around $K$-point with hexagonal symmetry. The FS topology of Ti$_2$AlC remains unchanged under pressure (10 GPa).

For Ti$_2$AlN, there is an electron-like ring with tinny gap at the center ($\Gamma$–$A$ direction) and this gap increases as well as the ring breaks up into 4 pieces at 4 GPa (not shown here) which is also visible at 10 GPa. The electron-like FS for band 31 is cylindrical (centered to the $\Gamma$–$A$ path). For band 32, electron-like sheet (with hexagonal symmetry) is seen around the $\Gamma$-point. Electron-like sheets appear around at the corner of the Brillouin zone for bands 33 and 34. There are little change observable in FS for bands 31 and 32 under 10 GPa. The FS topologies of Ti$_2$Al$X$ ($X$ = B, C, and N) at 0 GPa are in complete agreement with the previous studies [40,83].

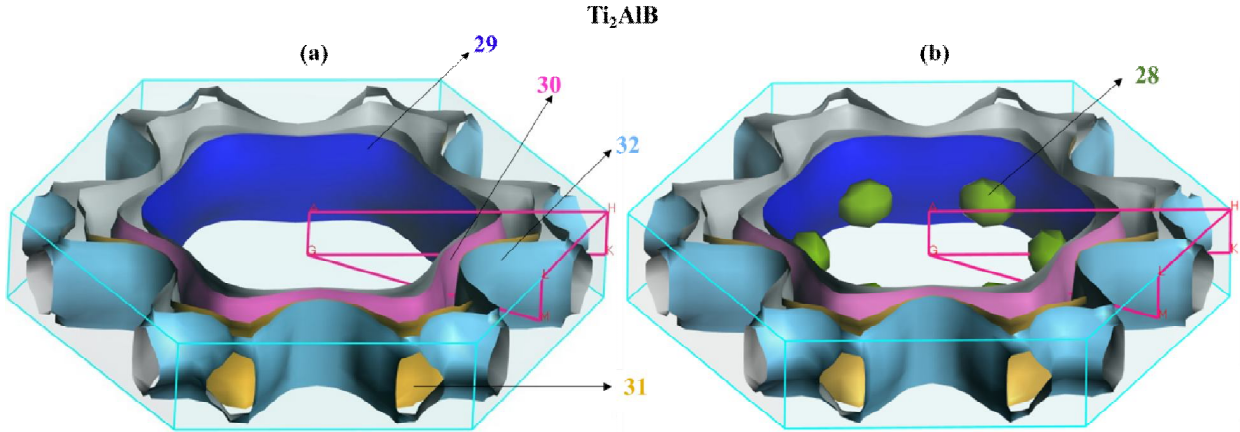

**Fig. 16.** The Fermi surfaces of Ti$_2$AlB at (a) 0 and (b) 10 GPa.

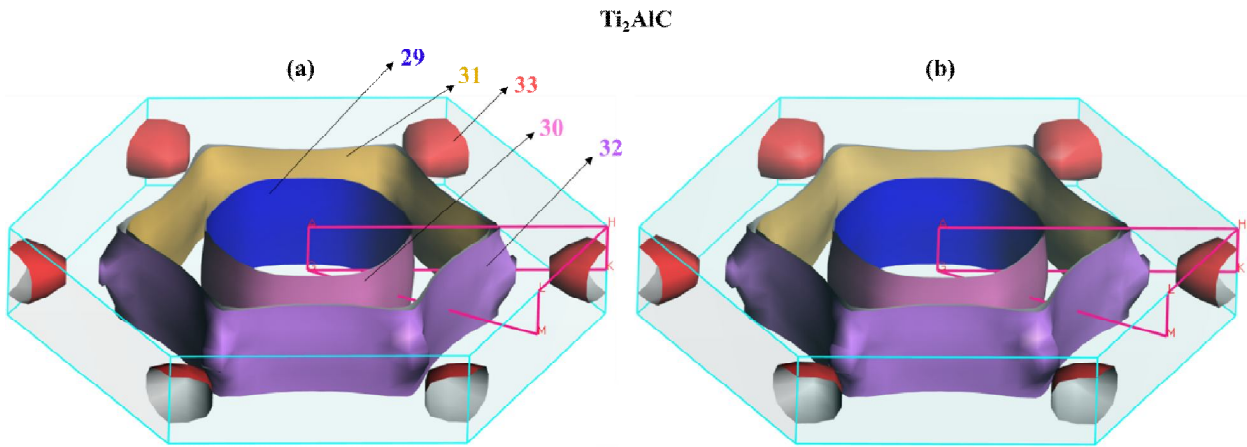

**Fig. 17.** The Fermi surfaces of Ti$_2$AlC at (a) 0 and (b) 10 GPa.

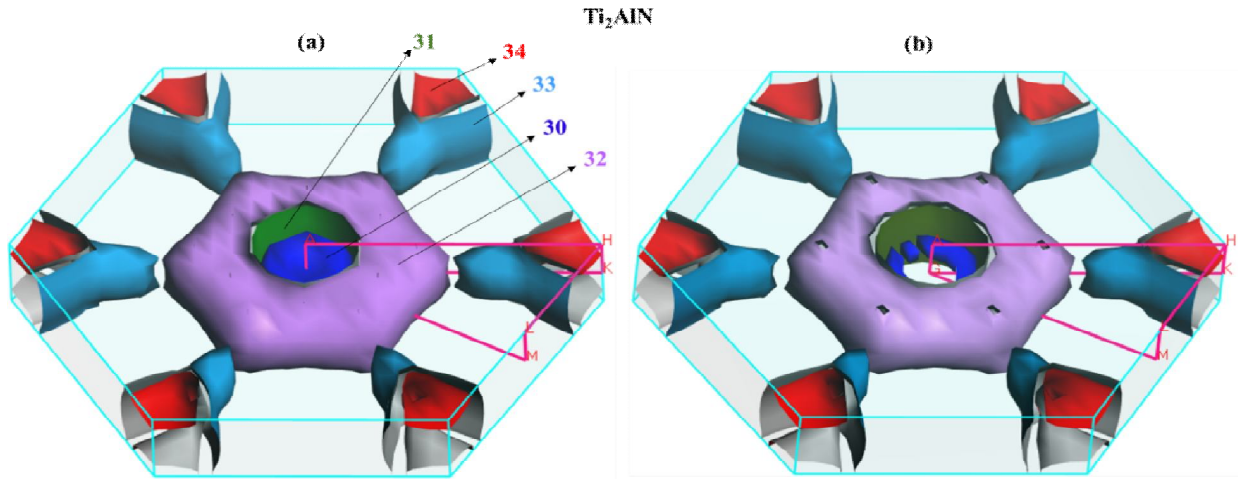

**Fig. 18.** The Fermi surfaces of Ti$_2$AlN at (a) 0 and (b) 10 GPa.

### 3.8. Optical properties

In optical properties, we study the response of a material to the irradiation of electromagnetic energy. Unlike semiconductor and insulator, metallic systems require the plasma frequency and Drude damping correction [84,85]. The calculation of the optical parameters of Ti$_2$AlX (X = B, C, and N) have done using a plasma energy of 5 eV and Drude damping of 0.05 eV with 0.5 eV energy smearing. The optical parameters of the compounds for electric field along the [100] polarization as a function of photon energy up to 25 eV under 0 and 10 GPa were studied to understand their optical response. Results are shown in Fig. 19-20.

The frequency dependent real ($\epsilon_1$) and imaginary ($\epsilon_2$) parts of the complex dielectric function in the energy range 0–25 eV for the compounds are depicted in Fig. 19a-b. For a material, $\varepsilon_1(\omega)$ elucidates the polarizability and light speed in the system, whereas $\varepsilon_2(\omega)$ is related to the band structure and absorptive behavior of the material [47]. The intraband transitions are dominated at low energy in metals and the interband transitions strongly correlates with the electronic band structure. The large negative values of $\epsilon_1$ for all three materials in the low energy region indicate their Drude-like behavior (Fig. 19a).

The investigation of complex refractive index of materials is crucial for designing optoelectronic devices. The refractive index $n(\omega)$ and extinction coefficient $k(\omega)$ of a material are incorporated into complex refractive index as: $N(\omega) = n(\omega) + ik(\omega)$. Fig. 19c and 19d represent $n(\omega)$ and $k(\omega)$ as a function of energy for Ti$_2$AlX MAX phases, respectively. The maximum of both $n(\omega)$ and $k(\omega)$ for all three cases is obtained at 0 eV and decreases with increasing energy. The density of accessible electronic states often drops as one moves further from the Fermi level, which causes the values of $n(\omega)$ and $k(\omega)$ to fall as energy increases. A compound with a larger value of $k(\omega)$ exhibits higher light attenuation ability. Refractive indices of all three MAX compounds are large in the visible region. Thus, these compounds can be used in optoelectronic display devices.

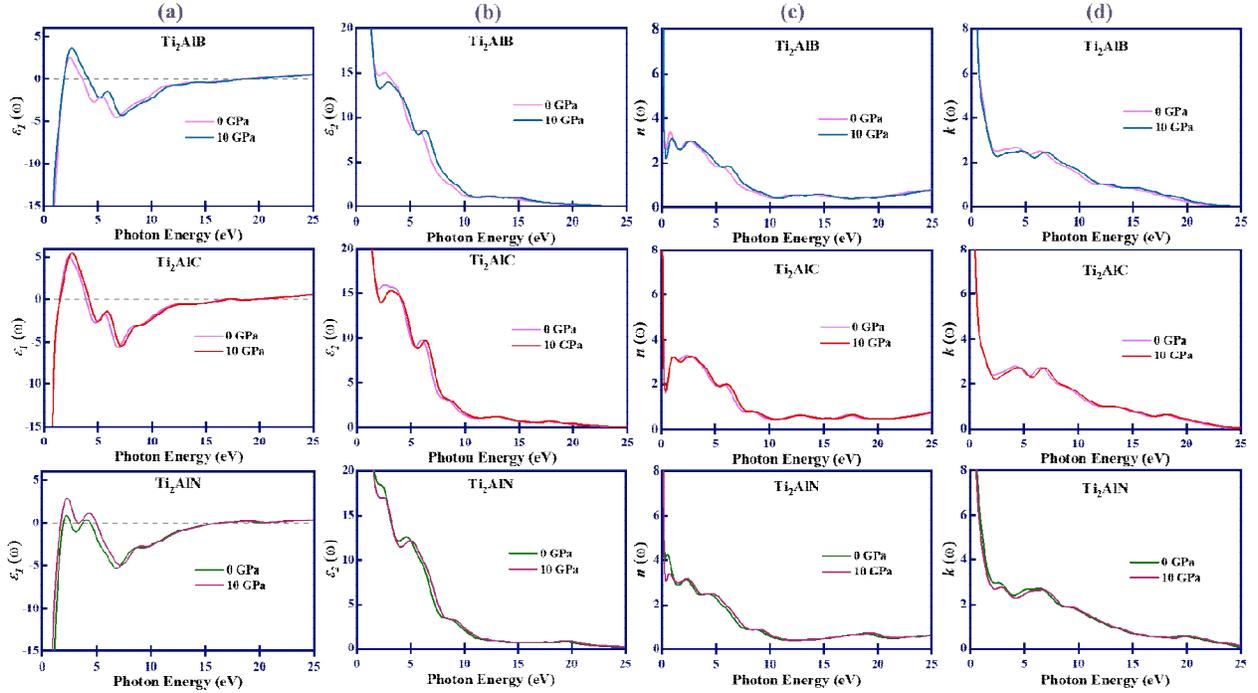

**Fig. 19.** Photon energy dependence of (a, b) dielectric constant ($\varepsilon_1$, and $\varepsilon_1$), and (c, d) refractive index (*n* and *k*) of Ti$_2$Al*X* (*X* = B, C, and N) at 0 and 10 GPa.

Optical reflectivity [$R(\omega)$] of the materials was studied and presented in Fig. 20a. At both 0 and 10 GPa, reflectivity is maximum at zero photon energy and decreases almost linearly around 2 eV for all three compounds. High *R* is caused by the interaction of incident light with free electrons in metals. $R_{max}$ at zero photon energy predicted shiny nature of the compounds. The value of *R* is almost nonselective and remains above 40% in the energy range from 0 to ~14 eV which implies that the studied MAX phase materials can be used as an efficient reflector of solar radiation. A material's electronic band structure is impacted by pressure. The density of states at the Fermi level varies with increasing pressure. These modifications lead to pressure dependent variation in the reflectivity.

The optical absorption [$\alpha(\omega)$] spectra of the compounds at 0 and 10 GPa are shown in Fig. 20b. The photon absorption for all the studied compounds begins at 0 eV indicating their metallic nature. This finding is also consistent with the band structure and DOS studies (see section 3.7.1). The absorption peaks for Ti$_2$AlB, Ti$_2$AlC, and Ti$_2$AlN are observed ~7 eV. Ti, Al, and the other elements (B, C, or N) contribute to the largest absorption peaks, which are connected to their electronic band structures. All three MAX compounds absorb UV radiation efficiently.

Fig. 20c displays the photoconductivity, $\sigma(\omega)$, of Ti$_2$Al*X* at 0 and 10 GPa. Photoconductivity initiates from zero photon energy for all studied compounds at both pressures, manifesting again the metallic behavior. This is also consistent with the electronic band structure, optical absorption, and dielectric function studies. Several features in the optical conductivity spectrum are related to the bulk plasmon excitations, and caused by valence to conduction band electron transition [37].

The electron energy-loss function [$L(\omega)$] spectra of the studied materials at 0 and 10 GPa are illustrated in Fig. 20d. The electron energy-loss function of a material describes the energy lost by a fast-moving electron traveling through the material. These interactions may include phonon excitation, interband and intraband transitions, plasmon excitations, inner shell ionizations and Cerenkov radiations [37]. The highest peak in energy-loss function spectrum for a material, associated with the plasma resonance, which appears at a specific photon frequency called plasma frequency ($\omega_p$) of the material [86]. At this specific energy, the charge carriers undergo collective oscillations whose frequency depends on the effective mass and concentration of the electrons. The highest peak of $L(\omega)$ for Ti$_2$AlB, Ti$_2$AlC, and Ti$_2$AlN at 0 GPa is about 18.49 eV, 21 eV, and 21.75 eV, respectively. A material becomes transparent and exhibits insulator-like optical features for the incident photon frequency higher than its $\omega_p$. The peak energy also indicates metallic to dielectric transition; above this point the material shows dielectric features [38]. The plasma frequency is given by the Drude model as $\omega_p = \sqrt{(n_e e^2)/(\varepsilon_0 m_e)}$, where $n_e$ is the electron density, $e$ is the elementary charge, $\varepsilon_0$ is the the permittivity of free space, and $m_e$ is the effective mass of the electrons. The basic feature of the connection between free electron density and collective electronic excitations in metals and other conductive materials is this relationship [84,85].

Overall variation in the optical parameters with pressure is weak (up to 10 GPa). Like many other MAX and MAB phase compounds [87-90], there is slight optical anisotropy with respect to electric field polarization direction observed in all three MAX compounds under investigation (not shown here).

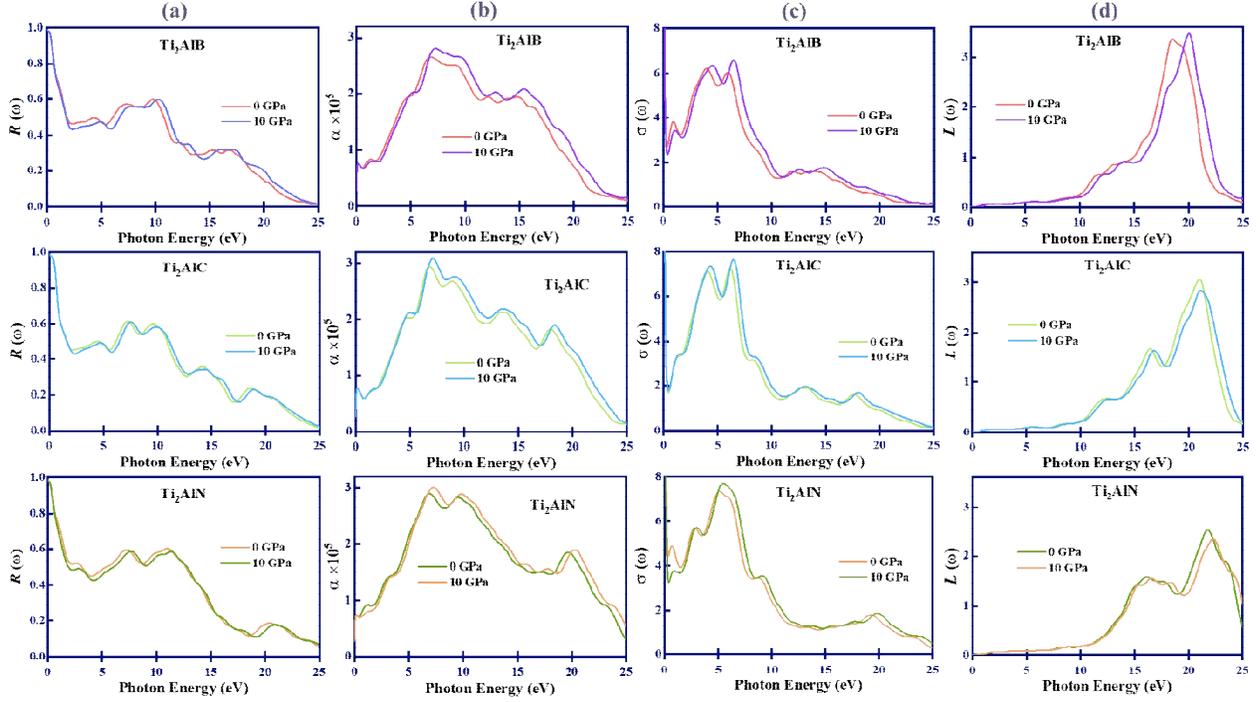

**Fig. 20.** Photon energy dependence of (a) $R(\omega)$, (b) $\alpha(\omega)$, (c) $\sigma(\omega)$, and (d) $L(\omega)$ of Ti$_2$Al$X$ ($X$ = B, C, and N) compounds at 0 and 10 GPa.

### 3.9. Theoretical Vickers hardness

The theoretical Vickers hardness ($H_V$) is an important mechanical parameter connected with bond overlap population and bond hardness between the atoms. The theoretical Vickers hardness of the MAX phases was acquired using following equations [91]:

$$H_V = \left[\prod^{\mu}(H_V^{\mu})^{n^{\mu}}\right]^{1/\sum n^{\mu}} \tag{9}$$

$$H_V^{\mu} = 740(P^{\mu} - P^{\mu'})(V_b^{\mu})^{-5/3} \tag{10}$$

$$V_b^{\mu} = \frac{(d^{\mu})^3}{\sum_V[(d^{\mu})^3 n^{\mu}]} \tag{11}$$

where, $d^{\mu}$ and $P^{\mu}$ define the Mulliken bond length and bond overlap population of the $\mu$-type bond, respectively. $P^{\mu'}(= n_{free}/V)$ is the metallic population ($n_{free}$ = number of free electrons), $n^{\mu}$ is the number of $\mu$-type bond per unit volume, $V_b^{\mu}$ is the bond volume of $\mu$-type bond, and $\sum n^{\mu}$ represents total number bonds in the unit cell. The constant 740 is a proportionality coefficient. The calculated intrinsic hardness values of Ti$_2$Al$X$ ($X$ = B, C, and N) are listed in Table 2.

The results indicate that all three compounds are relatively soft, with Ti$_2$AlC is comparatively harder among them. The value of $P^\mu$ for each bond between elements defines the degree of overlap of the electron clouds that form the bond. For all the compounds the longer and smaller values of $d^\mu$ and $P^\mu$ for Ti-Al bonds in comparison to Ti-$X$ bonds are a sign of lower bond hardness. This suggests that the bond breaking will start with Ti-Al under pressure or temperature.

Interestingly, the hardness of the materials decreases with pressure which contradicts the general trend of materials being harder under pressure. This comes from decrease in Ti-Al bond hardness under pressure, which results in a lower overall hardness. This decrease in mechanical hardness with pressure also correlates with Pugh's ratio and micro hardness (see Fig. 4). The calculated $n_{free}$ values for Ti$_2$AlB, Ti$_2$AlC, and Ti$_2$AlN at 0 GPa is 15.71, 9.37, and 1.89, respectively. The $n_{free}$ values for Ti$_2$AlB and Ti$_2$AlC are significantly higher than that of Ti$_2$AlN and increases up to10 GPa. As a consequence, the metallic bonding in the systems also increases under pressure.

**Table 2:** The calculated $\mu$-type bonds, number of $\mu$-type bond $n^\mu$, bond length $d^\mu$ (Å), bond volume $v_b^\mu$ (Å$^3$), bond overlap population $P^\mu$, metallic population $P^{\mu'}$, bond hardness $H_V^\mu$ (GPa), and hardness $H_V$ (GPa) of Ti$_2$Al$X$ ($X$ = B, C, and N) at 0 GPa and 10 GPa.

| Phases | Pressure | $\mu$-type bond | $n^\mu$ | $d^\mu$ | $v_b^\mu$ | $P^\mu$ | $P^{\mu'}$ | $H_V^\mu$ | $H_V$ |
|---|---|---|---|---|---|---|---|---|---|
| Ti$_2$AlB | 0 | Ti-B | 4 | 2.186 | 8.633 | 1.11 | 0.133 | 19.909 | 1.601 |
| | | Ti-Al | 4 | 2.939 | 20.985 | 0.60 | | 2.166 | |
| | 10 | Ti-B | 4 | 2.149 | 8.320 | 1.07 | 0.143 | 20.076 | 1.598 |
| | | Ti-Al | 4 | 2.838 | 19.172 | 0.54 | | 2.138 | |
| Ti$_2$AlC | 0 | Ti-C | 4 | 2.108 | 7.622 | 1.03 | 0.086 | 23.679 | 1.655 |
| | | Ti-Al | 4 | 2.896 | 19.772 | 0.55 | | 2.378 | |
| | 10 | Ti-C | 4 | 2.078 | 7.408 | 1.01 | 0.089 | 24.202 | 1.649 |
| | | Ti-Al | 4 | 2.812 | 18.367 | 0.48 | | 2.261 | |
| Ti$_2$AlN | 0 | Ti-N | 4 | 2.083 | 7.349 | 0.82 | 0.018 | 21.359 | 1.509 |
| | | Ti-Al | 4 | 2.837 | 18.561 | 0.24 | | 1.261 | |
| | 10 | Ti-N | 4 | 2.056 | 7.135 | 0.80 | 0.021 | 21.814 | 1.388 |
| | | Ti-Al | 4 | 2.767 | 17.401 | 0.12 | | 0.630 | |

## 4. Conclusions

We have employed DFT based first-principles calculations to conduct extensive studies of physical properties of Ti$_2$Al$X$ ($X$ = B, C, and N) MAX phases. The (hydrostatic) pressure-dependent structural, mechanical, thermophysical, phonon dynamical behavior, electronic (band, DOS, and FS topology), and optical properties are comprehensively explored for the very first time. The studied compounds possess structural, mechanical, and dynamical stability both at ambient and under pressure. The estimated structural parameters decrease with increasing pressure. Ti$_2$AlB undergoes brittle-ductile transition at about 8 GPa whereas Ti$_2$AlC and Ti$_2$AlN remain brittle up to 10 GPa. The compounds possess a moderate level of mechanical and optical anisotropy. The mixed bonding nature was found in Ti$_2$Al$X$ compounds. All three MAX phases are moderately machinable which increases gradually under pressure. The metallic nature of the compounds is confirmed from their electronic (band structure, DOS, and FS) and optical studies. The FS topologies are studied. A distinct electron-pocket with hexagonal symmetry in the BZ appears in Ti$_2$AlB under 10 GPa. Otherwise, the pressure effect on electronic band structure and FS is small that make the materials alluring for industrial applications under harsh and extreme conditions i.e., conditions involving large pressure changes. The estimated Coulomb pseudopotential implies electronic correlations in the order Ti$_2$AlB > Ti$_2$AlN > Ti$_2$AlC. The TDOS and Coulomb pseudopotential of Ti$_2$AlB increases with increasing pressure, while Ti$_2$AlC and Ti$_2$AlN show reverse response. The charge density difference plots reveal direction dependency of atomic bonding. At room temperature, the phonon thermal conductivity of Ti$_2$AlC is significantly higher than other two. All three compounds have some potential as thermal insulation materials [92,93]. It is predicted that Ti$_2$AlC and Ti$_2$AlN have much higher Debye and melting temperature than Ti$_2$AlB. The optical properties suggest their suitability in some optical applications.

In summary, we have investigated systematically a large number of physical properties of the Ti$_2$Al$X$ ($X$ = B, C, and N) MAX phases in this paper. The studied compounds have considerable potential in thermo-mechanical and optoelectronic applications. We hope that these results will stimulate researchers to investigate these MAX phase nanolaminates in greater detail, both theoretically as well as experimentally in near future.

## Acknowledgements
Authors are grateful to the Department of Physics, Chittagong University of Engineering & Technology (CUET), Chattogram-4349, Bangladesh, for providing the computing facilities for this work. This work was also carried out with the aid of a grant (grant number: 21-378 RG/PHYS/AS_G-FR3240319526) from UNESCO-TWAS and the Swedish International Development Cooperation Agency (SIDA). The views expressed herein do not necessarily represent those of UNESCO-TWAS, SIDA, or its Board of Governors.

## Data availability
The data sets generated and/or analyzed in this study are available from the corresponding author on reasonable request.

## Author Contributions
**M. I. Naher:** Methodology, Formal analysis, Writing − original draft; **M. Montasir:** Formal analysis, Writing – original draft; **M. Y. H. Khan:** Formal analysis, Writing – original draft;  **M. A. Ali:** Formal analysis, Writing − original draft, Validation**; M. M. Hossain:** Formal analysis, Writing review & editing, Validation; **M. M. Uddin:** Writing review & editing, Validation; **M. Z.**

**Hasan:** Writing review & editing, Validation; **M. A. Hadi:** Writing review & editing, Validation; **S. H. Naqib:** Conceptualization, Formal analysis, Writing review & editing, Validation, Supervision.

## Competing Interests

The authors declare that they have no known competing financial interests or personal relationships that could have appeared to influence the work reported in this paper.

## References

1. Rackl T, Johrendt D. The MAX phase borides $Zr_2SB$ and $Hf_2SB$. Solid State Sciences 2020;106:106316. https://doi.org/10.1016/j.solidstatesciences.2020.106316
2. Zhang Z, Duan X, Jia D, Zhou Y, Zwaag SVD. On the formation mechanisms and properties of MAX phases: A review. Journal of the European Ceramic Society 2021;41:3851. https://doi.org/10.1016/j.jeurceramsoc.2021.02.002
3. Gonzalez-Julian J. Processing of MAX phases: from synthesis to applications. J. Am. Ceram. Soc. 2021;104:659. DOI: 10.1111/jace.17544
4. Rana MR, Islam S, Hoque K, Biswas GG, Hossain ME, Naqib SH, Ali MA. DFT prediction of the stability and physical properties of $M_2GaB$ (M = Sc, V, Nb, Ta). Journal of Materials Research and Technology 2023;24:7795. https://doi.org/10.1016/j.jmrt.2023.05.008
5. Barsoum MW, Ali M, El-Raghy T. Processing and characterization of $Ti_2AlC$, $Ti_2AlN$, and $Ti_2AlC_{0.5}N_{0.5}$. Metall. Mater. Trans. A 2000;31:1857. https://doi.org/10.1007/s11661-006-0243-3
6. Barsoum MW, Brodkin D, El-Raghy T. Layered machinable ceramics for high temperature applications. Scr. Mater. 1997;36:535. https://doi.org/10.1016/S1359-6462(96)00418-6
7. Barsoum MW. The $M_{N+1}AX_N$ phases: A new class of solids: Thermodynamically stable nanolaminates. Progress in Solid State Chemistry 2000;28:201. https://doi.org/10.1016/S0079-6786(00)00006-6
8. Bai Y, Srikanth N, Chua CK, Zhou K. Density functional theory study of $M_{n+1}AX_n$ phases: A review. Critical Reviews in Solid State and Materials Sciences 2019;44:56. https://doi.org/10.1080/10408436.2017.1370577
9. Surucu G. Investigation of structural, electronic, anisotropic elastic, and lattice dynamical properties of MAX phases borides: An ab-initio study on hypothetical $M_2AB$ (M = Ti, Zr, Hf; A = Al, Ga, In) compounds. Materials Chemistry and Physics 2018;203:106. https://doi.org/10.1016/j.matchemphys.2017.09.050
10. Ali MA, Qureshi MW. Newly synthesized MAX phase $Zr_2SeC$: DFT insights into physical properties towards possible applications. RSC Adv. 2021;11:16892. DOI: 10.1039/D1RA02345D
11. Khazaei M, Arai M, Sasaki T, Estili M, Sakka Y. Trends in electronic structures and structural properties of MAX phases: a first-principles study on $M_2AlC$ (M = Sc, Ti, Cr, Zr, Nb, Mo, Hf, or Ta), $M_2AlN$, and hypothetical $M_2AlB$ phases. J. Phys.: Condens. Matter 2014;26:505503. DOI: 10.1088/0953-8984/26/50/505503

12. Sun ZM. Progress in research and development on MAX phases: a family of layered ternary compounds. Int. Mater. Rev. 2011;56(3):143. DOI: 10.1179/1743280410Y.0000000001
13. Hettinger JD, Lofland SE, Finke P, *et al.* Electrical transport, thermal transport, and elastic properties of $M_2$AlC ($M$ = Ti, Cr, Nb, and V). Phys. Rev. B 2005;72:115120. https://doi.org/10.1103/PhysRevB.72.115120
14. Du YL, Sun ZM, Hashimoto H. First-principles study on phase stability and compression behavior of $Ti_2$SC and $Ti_2$AlC. Physica B 2010;405:720. https://doi.org/10.1016/j.physb.2009.09.093
15. Suna Z, Li S, Ahuja R, Schneider JM. Calculated elastic properties of $M_2$AlC (M = Ti, V, Cr, Nb and Ta). Solid State Communications 2004;129:589. https://doi.org/10.1016/j.ssc.2003.12.008
16. Cover MF, Warschkow O, Bilek MMM, McKenzie DR. A comprehensive survey of $M_2$AX phase elastic properties. J. Phys.: Condens. Matter 2009;21:305403. DOI: 10.1088/0953-8984/21/30/305403
17. Persson POÅ, Rosén J, McKenzie DR, Bilek MMM. Formation of the *MAX*-phase oxycarbide $Ti_2AlC_{1-x}O_x$ studied via electron energy-loss spectroscopy and first-principles calculations. Phys. Rev. B 2009;80:092102. https://doi.org/10.1103/PhysRevB.80.092102
18. Wang XH, Zhou YC. High-Temperature oxidation behavior of $Ti_2$AlC in air. Oxidation of Metals 2003;59:303. https://doi.org/10.1023/A:1023092027697
19. Low IM. An overview of parameters controlling the decomposition and degradation of Ti-based $M_{n+1}AX_n$ phases. Materials 2019;12(3):473. https://doi.org/10.3390/ma12030473
20. Hug G, Fries E. Full-potential electronic structure of $Ti_2$AlC and $Ti_2$AlN. Phys. Rev. B 2002;65:113104. https://doi.org/10.1103/PhysRevB.65.113104
21. Dhakal C, Aryal S, Sakidja R, Ching WY. Approximate lattice thermal conductivity of MAX phases at high temperature. Journal of the European Ceramic Society 2015;35:3203. https://doi.org/10.1016/j.jeurceramsoc.2015.04.013
22. Bai Y, He X, Wang R. Lattice dynamics of Al-containing MAX-phase carbides: a first-principles study. J. Raman Spectrosc. 2015;46:784. https://doi.org/10.1002/jrs.4720
23. Qureshi MW, Ma X, Tang G, Paudel R. Ab initio predictions of structure and physical properties of the $Zr_2$GaC and $Hf_2$GaC MAX phases under pressure. Sci Rep 2021;11:3260. https://doi.org/10.1038/s41598-021-82402-1
24. Fakhera F, Hossain K, Khanom S, Hossain MK, Ahmed F. Effect of pressure in tuning the physical properties of MAX phase $Zr_2$AN (A = In, Ga): A DFT scheme. Results in Physics 2023;53:106912. https://doi.org/10.1016/j.rinp.2023.106912
25. Hossain MS, Jahan N, Hossain MM, Uddin MM, Ali MA, High pressure mediated physical properties of $Hf_2$AB (A = Pb, Bi) via DFT calculations. Materials Today Communications 2023;34:105147. https://doi.org/10.1016/j.mtcomm.2022.105147
26. Clark SJ, Segall MD, Pickard CJ, *et al.* First principles methods using CASTEP. Z. Kristallogr 2005;220:567. https://doi.org/10.1524/zkri.220.5.567.65075.

27. Parr RG. Density functional theory. Ann Rev Phys Chern 1983;34(1):631. https://doi.org/10.1146/annurev.pc.34.100183.003215
28. Materials studio CASTEP manual © Accelrys2010. http://www.tcm.phy.cam.ac.uk/castep/documentation/WebHelp/CASTEP.html.
29. Perdew JP, *et al.* Restoring the density-gradient expansion for exchange in solids and surfaces. Phys. Rev. Lett. 2008;100:136406. https://doi.org/10.1103/PhysRevLett.100.136406
30. Vanderbilt D. Soft self-consistent pseudopotentials in a generalized eigenvalue formalism. Phys. Rev. B 1990;41:7892. https://doi.org/10.1103/PhysRevB.41.7892
31. Packwood D, *et al.* A universal preconditioner for simulating condensed phase materials. J. Chem. Phys. 2016;144:164109. https://doi.org/10.1063/1.4947024
32. Monkhorst HJ, Pack JD. Special points for Brillouin-zone integrations. Phys. Rev. B 1976;13:5188. https://doi.org/10.1103/PhysRevB.13.5188
33. Koukaras EN, Kalosakas G, Galiotis C, Papagelis K. Phonon properties of graphene derived from molecular dynamics simulations. Sci Rep 2015;5:12923. https://doi.org/10.1038/srep12923
34. Kresse G, Furthmüller J, Hafner J. Ab initio force constant approach to phonon dispersion relations of diamond and graphite. Europhysics Letters 1995;32:729. DOI: 10.1209/0295-5075/32/9/005
35. Nielsen OH, Martin RM. First-principles calculation of stress. Phys. Rev. Lett. 1983;50:697. https://doi.org/10.1103/PhysRevLett.50.697
36. Marius G. Kramers-Kronig relations (The Physics of Semiconductors), Springer US, Berlin Heidelberg, 2010.
37. Naderizadeh S, *et al.* Electronic and optical properties of Full-Heusler alloy $Fe_{3-x}Mn_xSi$. Eur. Phys. J. B 2012;85:144. https://doi.org/10.1140/epjb/e2012-20919-3
38. Sun J, Wang HT, He JL, Tian YJ. Ab initio investigations of optical properties of the high-pressure phases of ZnO. Phys. Rev. B 2005;71:125132. https://doi.org/10.1103/PhysRevB.71.125132
39. Schuster JC, Nowotny H, Vaccaro C. The ternary systems: Cr-Al-C, V-Al-C, and Ti-Al-C and the behavior of H-phases ($M_2$AlC). Journal of Solid State Chemistry 1980;32:213. https://doi.org/10.1016/0022-4596(80)90569-1
40. Karaca E, Byrne PJP, Hasnip PJ, Probert MIJ. Prediction of phonon‐mediated superconductivity in new Ti‐based $M_2AX$ phases. Sci Rep 2022;12:13198. https://doi.org/10.1038/s41598-022-17539-8
41. Basu S, Barsoum MW, Kalidindi SR. Sapphire: A kinking nonlinear elastic solid. Journal of Applied Physics 2006;99:063501. https://doi.org/10.1063/1.2179974
42. Poon B, Ponsonn L, Zhao J, Ravichandran G. Damage accumulation and hysteretic behavior of MAX phase materials. Journal of the Mechanics and Physics of Solids 2011;59:2238. https://doi.org/10.1016/j.jmps.2011.03.012

43. Naher MI, Naqib SH. An ab-initio study on structural, elastic, electronic, bonding, thermal, and optical properties of topological Weyl semimetal Ta$X$ ($X$ = P, As). Sci Rep 2021;11:5592. https://doi.org/10.1038/s41598-021-85074-z
44. Naher MI, Ali MA, Hossain MM, Uddin MM, Naqib SH. A comprehensive ab-initio insights into the pressure dependent mechanical, phonon, bonding, electronic, optical, and thermal properties of CsV$_3$Sb$_5$ Kagome compound. Results in Physics 2023;51:106742. https://doi.org/10.1016/j.rinp.2023.106742
45. Naher MI, Naqib SH. First-principles insights into the mechanical, optoelectronic, thermophysical, and lattice dynamical properties of binary topological semimetal BaGa$_2$. Results in Physics 2022;37:105507. https://doi.org/10.1016/j.rinp.2022.105507
46. Born M, Hang K. Dynamical theory and experiments I. Berlin: Springer-Verlag Publishers; 1982.
47. Naher MI, Naqib SH. Possible applications of Mo$_2$C in the orthorhombic and hexagonal phases explored via ab-initio investigations of elastic, bonding, optoelectronic and thermophysical properties. Results in Physics 2022;37:105505. https://doi.org/10.1016/j.rinp.2022.105505
48. Naher MI, Afzal MA, Naqib SH. A comprehensive DFT based insights into the physical properties of tetragonal superconducting Mo$_5$PB$_2$. Results in Physics 2021;28:104612. https://doi.org/10.1016/j.rinp.2021.104612
49. Voigt W. Lehrbuch der Kristallphysik. Leipzig: Taubner; 1928. p. 962.
50. Reuss A. Berechnung der fließgrenze von mischkristallen auf grund der plastizitätsbedingung für einkristalle. Z Angew Math Mech 1929;9:49. https://doi.org/10.1002/zamm.19290090104
51. Hill R. The elastic behaviour of a crystalline aggregate. Proc. Phys. Soc. A 1952;65:349. DOI: 10.1088/0370-1298/65/5/307
52. Wang X, Xiang H, Sun X, Liu J, Hou F, Zhou Y. Mechanical properties and damage tolerance of bulk Yb$_3$Al$_5$O$_{12}$ ceramic. Journal of Materials Science & Technology 2015;31:369. https://doi.org/10.1016/j.jmst.2015.01.002
53. Ali MA, Hossain MM, Uddin MM, Hossain MA, Islam AKMA, Naqib SH. Physical properties of new MAX phase borides M$_2$SB (M = Zr, Hf and Nb) in comparison with conventional MAX phase carbides M$_2$SC (M = Zr, Hf and Nb): Comprehensive insights. Journal of Materials Research and Technology 2021;11:1000. https://doi.org/10.1016/j.jmrt.2021.01.068
54. Jong MD, Winter I, Chrzan DC, Asta M. Ideal strength and ductility in metals from second- and third-order elastic constants. Phys. Rev. B 2017;96:014105. https://doi.org/10.1103/PhysRevB.96.014105
55. Greaves GN, Greer AL, Lakes RS, Rouxel T. Poisson's ratio and modern materials. Nat Mater 2011;10:823. https://doi.org/10.1038/nmat3134
56. Pan Y, Zheng WT, Guan WM, Zhang KH, Fan XF. First-principles study on the structure, elastic properties, hardness and electronic structure of $TM$B$_4$ ($TM$ = Cr, Re, Ru and Os) compounds. Journal of Solid State Chemistry 2013;207:29. https://doi.org/10.1016/j.jssc.2013.09.012

57. Slack GA. The thermal conductivity of nonmetallic crystals. Solid state Physics 1979;34:1. https://doi.org/10.1016/S0081-1947(08)60359-8
58. Meneve J, Vercammen K, Dekempeneer E, Smeets J. Thin tribological coatings: magic or design? Surf Coat Technol 1997;94–95:476. https://doi.org/10.1016/S0257-8972(97)00430-1
59. Cheng YT, Page T, Pharr GM, Swain M, Wahl KJ. (Eds.), Fundamentals and applications of instrumented indentation in multidisciplinary research. Journal of Materials Research 2004;19:1.
60. Donnet C, Erdemir A. Solid lubricant coatings: recent developments and future trends. Tribol Lett 2004;17(3):389. https://doi.org/10.1023/B:TRIL.0000044487.32514.1d
61. Chen XQ, Niu H, Li D, Li Y. Modeling hardness of polycrystalline materials and bulk metallic glasses. Intermetallics 2011;19:1275. https://doi.org/10.1016/j.intermet.2011.03.026
62. Pettifor DG. Theoretical predictions of structure and related properties of intermetallics. Mater Sci Technol 1992;8(4):345. DOI: 10.1179/mst.1992.8.4.345
63. Zeng X, Peng R, Yu Y, Hu Z, Wen Y, Song L. Pressure effect on elastic constants and related properties of $Ti_3Al$ intermetallic compound: a first-principles study. Materials 2018;11:2015. https://doi.org/10.3390/ma11102015
64. Qu D, Li C, Bao L, Kong Z, Duan Y. Structural, electronic, and elastic properties of orthorhombic, hexagonal, and cubic $Cu_3Sn$ intermetallic compounds in Sn–Cu lead-free solder. J Phys Chem Solids 2020;138:109253. https://doi.org/10.1016/j.jpcs.2019.109253
65. Ranganathan SI, Ostoja-Starzewski M. Universal elastic anisotropy index. Phys. Rev. Lett. 2008;101:055504. https://doi.org/10.1103/PhysRevLett.101.055504
66. Gaillac R, Pullumbi P, Coudert FX. ELATE: an open-source online application for analysis and visualization of elastic tensors. J Phys Condens Matter 2016;28(27):275201. DOI: 10.1088/0953-8984/28/27/275201
67. Anderson OL. A simplified method for calculating the Debye temperature from elastic constants. J Phys Chem Solids 1963;24(7):909. https://doi.org/10.1016/0022-3697(63)90067-2
68. Naher MI, Parvin F, Islam AKMA, Naqib SH. Physical properties of niobium-based intermetallics ($Nb_3B$; $B$ = Os, Pt, Au): a DFT-based ab-initio study. Eur Phys J B 2018;91:289. https://doi.org/10.1140/epjb/e2018-90388-9
69. Cheng L, Yan QB, Hu M. The role of phonon–phonon and electron–phonon scattering in thermal transport in $PdCoO_2$. Phys. Chem. Chem. Phys. 2017;19:21714. DOI: 10.1039/C7CP03667A
70. Fine ME, Brown LD, Marcus HL. Elastic constants versus melting temperature in metals. Scr Metall 1984;18:951. https://doi.org/10.1016/0036-9748(84)90267-9
71. Morelli DT, Slack GA. High lattice thermal conductivity solids. In: Shindé SL, Goela JS, editors. High Thermal Conductivity Materials. New York, NY: Springer; 2006. https://doi.org/10.1007/0-387-25100-6_2
72. Julian CL. Theory of heat conduction in rare-gas crystals. Phys. Rev. 1965;137:A128. https://doi.org/10.1103/PhysRev.137.A128

73. Clarke DR. Materials selection guidelines for low thermal conductivity thermal barrier coatings. Surf Coat Technol 2003;163-164:67. https://doi.org/10.1016/S0257-8972(02)00593-5
74. Cahill DG, Pohl RO. Lattice vibrations and heat transport in crystals and glasses. Annu. Rev. Phys. Chem. 1988;39:93.
75. Shafique A, Shin YH. Ultrahigh and anisotropic thermal transport in the hybridized monolayer ($BC_2N$) of boron nitride and graphene: a first-principles study. Phys. Chem. Chem. Phys. 2019;21:17306. DOI: 10.1039/C9CP02068C
76. Whalley LD, Frost JM, Morgan BJ, Walsh A. Impact of nonparabolic electronic band structure on the optical and transport properties of photovoltaic materials. Phys. Rev. B 2019;99:085207. https://doi.org/10.1103/PhysRevB.99.085207
77. Schlüter M. Effect of high pressure on the electronic density of states of aluminium. phys. stat. sol. (b) 1971;43:141. https://doi.org/10.1002/pssb.2220430115
78. Kostrzewa M, Szczęśniak R, Kalaga JK, Wrona IA. Anomalously high value of Coulomb pseudopotential for the $H_5S_2$ superconductor. Sci Rep 2018;8:11957. https://doi.org/10.1038/s41598-018-30391-z
79. Bennemann KH, Garland JW. Theory for Superconductivity in *d* band metals. AIP Conf. Proc. 1972;4:103. https://doi.org/10.1063/1.2946179
80. Hadi MA, Naqib SH, Christopoulos SRG, Chroneos A, Islam AKMA. Mechanical behavior, bonding nature and defect processes of $Mo_2ScAlC_2$: A new ordered MAX phase. Journal of Alloys and Compounds 2017;724:1167. https://doi.org/10.1016/j.jallcom.2017.07.110
81. Chowdhury A, Ali MA, Hossain MM, Uddin MM, Naqib SH, Islam AKMA. Predicted MAX phase $Sc_2InC$: Dynamical stability, vibrational and optical properties. Phys. Status Solidi B 2017;255:1700235. https://doi.org/10.1002/pssb.201700235
82. Cudazzo P, *et al.* Ab Initio description of high-temperature superconductivity in dense molecular hydrogen. Phys. Rev. Lett. 2008;100:257001. https://doi.org/10.1103/PhysRevLett.100.257001
83. Mauchamp V, *et al.* Anisotropy of the resistivity and charge-carrier sign in nanolaminated $Ti_2AlC$: Experiment and ab initio calculations. Phys. Rev. B 2013;87:235105. https://doi.org/10.1103/PhysRevB.87.235105
84. Li S, Ahuja R, Barsoum MW, Jena P, Johansson B. Optical properties of $Ti_3SiC_2$ and $Ti_4AlN_3$. Appl. Phys. Lett. 2008;92:221907. https://doi.org/10.1063/1.2938862
85. Wang H, Chen Y, Kaneta Y, Iwata S. First-principles study on effective doping to improve the optical properties in spinel nitrides. Journal of Alloys and Compounds 2010;491:550. https://doi.org/10.1016/j.jallcom.2009.10.267
86. Fox M. Optical properties of solids. New York: Academic Press; 1972
87. Ali MA, Hadi MA, Hossain MM, Naqib SH, Islam AKMA. Theoretical investigation of structural, elastic, and electronic properties of ternary boride MoAlB. Phys. Status Solidi B 2017;254:1700010. https://doi.org/10.1002/pssb.201700010

88. Ali MA, Hossain MM, Islam AKMA, Naqib SH. Ternary boride $Hf_3PB_4$: Insights into the physical properties of the hardest possible boride MAX phase. Journal of Alloys and Compounds 2021;857:158264. https://doi.org/10.1016/j.jallcom.2020.158264
89. Hadi MA, Roknuzzaman M, Nasir MT, Monira U, Naqib SH, Chroneos A, Islam AKMA, Alarco JA, Ostrikov K. Effects of Al substitution by Si in $Ti_3AlC_2$ nanolaminate. Sci Rep 2021;11:3410. https://doi.org/10.1038/s41598-021-81346-w
90. Aktar MB, Parvin F, Islam AKMA, Naqib SH, Structural, elastic, electronic, bonding, thermo-mechanical and optical properties of predicted NbAlB MAB phase in comparison to MoAlB: DFT based ab-initio insights. Results in Physics 2023;52:106921. https://doi.org/10.1016/j.rinp.2023.106921
91. Gao FM. Theoretical model of intrinsic hardness. Phys. Rev. B 2006;73:132104. https://doi.org/10.1103/PhysRevB.73.132104
92. Barsoum MW, Salama I, El-Raghy T, *et al*. Thermal and electrical properties of $Nb_2AlC$, (Ti, $Nb)_2AlC$ and $Ti_2AlC$. Metallurgical and Materials Transactions A 2002;33:2775. https://doi.org/10.1007/s11661-002-0262-7
93. Yue SY, Qin G, Zhang X, Sheng X, Su G, Hu M. Thermal transport in novel carbon allotropes with $sp^2$ or $sp^3$ hybridization: An ab initio study. Phys. Rev. B 2017;95:085207. https://doi.org/10.1103/PhysRevB.95.085207